\documentclass[bibyear]{aa}  
\usepackage{graphicx}
\usepackage{multirow}
\usepackage{natbib}
\usepackage{xcolor}
\usepackage{lscape}
\usepackage{placeins}
\usepackage{seqsplit}
\newcommand{\code}[1]{\textsc{#1}}
\usepackage{txfonts}
\usepackage[]{hyperref}
\usepackage{enumitem}

\begin{document} 
   \title{Radio emission from flaring stars and brown dwarfs}
   \author{K. Perger\inst{1,2}         
          \and
         B. Seli\inst{1,2}
           \and
         K. Vida\inst{1,2}}
         \institute{
             Konkoly Observatory, HUN-REN Research Centre for Astronomy and Earth Sciences, Konkoly Thege Mikl\'os \'ut 15-17, 1121 Budapest, Hungary
    \and    CSFK, MTA Centre of Excellence, Konkoly Thege Mikl\'os \'ut 15-17, 1121 Budapest, Hungary\\ \email{perger.krisztina@csfk.org}}
   \date{Received April 4, 2025; Accepted June 24, 2025}
 
\abstract{The vicinities of intermediate-to-late type dwarf stars are considered as an adequate terrain to host planets suitable for life to form. However, they oftentimes show increased stellar activity, which should be taken into consideration when seeking potential habitable planetary systems.}{With the aim to reveal the effects of the magnetic field to the multi-band activity of dwarf stars, we search for associated radio emission for an extensive list of $14\,915$ brown dwarfs and $15\,124$ flaring stars.}{We utilised the first and second epoch catalogues and radio maps from all three epochs of the VLASS, supplemented with X-ray catalogues based on observations by the  \textit{ROSAT}, \textit{eROSITA}, and \textit{XMM-Newton} space telescopes, and 2-minute cadence optical light curves from the \textit{TESS} mission. The radio-detected sub-sample was queried for concurrent \textit{TESS} observations, and sources with coinciding light-curves were studied individually.}{We found no associated radio emission for brown dwarfs, and found $55$ radio counterparts for the sample of flaring stars, out of which seven have coincident \textit{TESS} observations. The radio-detected sample follows both the radio--X-ray and the period--activity relations. We found a strong correlation between the radio powers and the stellar parameters of surface gravity, radius, and mass. We found no connection between the flare rate and the radio variability. For radio-detected stars with available effective temperatures and rotational periods, we estimated gyrochronological ages,  which resulted in values of $T_\mathrm{gyro}\lesssim1$~Gyr, with the majority of the sample being younger than $150$~Myr. We found no strong connection between the occurrence of optical flares and radio variability for the individually studied stars.}{We conclude that radio emission from intermediate-to-late type flaring stars is of synchrotron nature, and shares a common origin with X-ray processes. It is created by a predominantly young stellar population, and is the collective contribution of stellar flares, accretion, and coronal heating.}


\maketitle

\section{Introduction}
Investigating the radio emission of intermediate or late-type dwarf stars provides a valuable tool for probing the magnetic activity, the properties of stellar coronae, and potential star--planet interactions, advancing our understanding of stellar magnetospheres and exoplanetary magnetic environments. Radio-emitting stars have been identified across the entire Hertzsprung–Russell diagram. The sources of the radio emission were identified as young stellar objects \citep[e.g.][]{2018A&ARv..26....3A,2019MNRAS.486.3664O,2024MNRAS.533.3862O}, T Tauri stars \citep[e.g.][]{2014ApJ...792L..18A,2018AJ....155...47A}, ultra-cool dwarfs \citep[e.g.][]{2015Natur.523..568H,2023Natur.619..272K,2023ApJ...951L..43R,2023A&A...675L...6V}, RS~CVn stars \citep[e.g.][]{1985A&A...149..343K,2021A&A...654A..21T}, eclipsing binary systems \citep[e.g.][]{1998A&A...329.1010U,2024ApJ...975...34F}, and footprints of star--planet interactions \citep[e.g.][]{2018MNRAS.481..217T,
2020NatAs...4..577V,2021NatAs...5.1233C,2021A&A...650L..20D,2021MNRAS.504.1511K,2021ApJ...919L..10P}. The detected radio emission can originate from both thermal and non-thermal processes, such as stellar magnetic activity, shock interactions, jets, and stellar pulsations \citep[e.g.][]{2002ARA&A..40..217G,2019PASP..131a6001M}. Unfortunately, attempts to directly detect thermal radio emission originating from cool stars were not successful yet \citep[e.g.][]{1997A&A...319..578V,2017A&A...599A.127F}. 
A wide range of physical processes is considered as the origin of the radio emission in stellar systems. One of the most common mechanisms is believed to originate from electrons moving within the stellar magnetosphere, producing non-thermal continuum emission via cyclotron, gyro-synchrotron, or synchrotron emission, depending on the velocity of the emitting particles \citep[e.g.][]{1985ARA&A..23..169D,1987ApJ...322..902D,1992ApJ...393..341L,2000A&A...362..281T,2007A&A...475..359G}. Additionally, stellar flares are described to release the built-up magnetic energy through the accelerated electrons into the stellar corona \citep[e.g.][]{2002ARA&A..40..217G}. This results in the heating of the corona, which often fuels the launch of the  associated X-ray emission in flaring stars \citep[e.g.][]{1990SoPh..130..265B,2002ARA&A..40..217G,2017MNRAS.465L..74W}. This is imprinted in the empirical correlation found between the (quiescent) radio- and X-ray emission \cite[e.g.][]{1993ApJ...405L..63G,1994A&A...285..621B}, hinting at a common origin of the two processes as the same population of particles interacting with the magnetic field. Radio emission of coronal origin is released in the form of type II (long-duration) radio bursts, as charged particles are affected by magnetic suppression and shocks while moving outwards through the stellar corona \citep[e.g.][and references therein]{2017PhDT.........8V}. Type II bursts are believed to be adequate tracers to coronal mass ejections,  occasional accomplices of stellar flares, which could greatly affect the potential habitability of the planetary system of the star (see \citealt{2024Univ...10..313V} and references therein). 
\citet{2022ApJ...926L..30V} discussed the possibility of low- and intermediate frequency radio emission of ultra-cool dwarfs sharing the same intrinsic mechanism, and whether that emission process is inherently coherent or incoherent in nature. Radio emission was also detected in the form coherent bursts during a large optical flare of Proxima Centauri, and was identified as a type IV burst, indicating the presence of coronal mass ejections on the nearby active M-dwarf \citep{2020ApJ...905...23Z}. In this paper we aim to study the radio emission properties of intermediate-to-late type stars, focusing on flaring stars and brown dwarfs.

\section{Sample selection and radio associations in the vicinity of the dwarf stars}

We defined two separate samples for the study, one consisting of brown dwarf candidates (BD), and one listing stars with recorded flaring activity (FS).
The sample of brown dwarf candidates was comprised of a data set containing $14\,915$ ultra-cool dwarf candidates, selected from a Gaia DR2 based catalogue of \citet{2018A&A...619L...8R} with spectral types between M7.5--L4. We used the catalogue of \cite{2025A&A...694A.161S} to compile the sample of flaring stars. This catalogue was created from 2-min cadence \textit{Transiting Exoplanet Survey Satellite (TESS)} light curves up to sector 69, using a combination of a deep neural network \citep[\code{flatwrm2},][]{flatwrm2} and manual vetting.
\code{flatwrm2} is a deep learning algorithm  using a recurrent neural network (RNN), based on long short-term memory (LSTM) layers. The network was retrained specifically for \textit{TESS} observations to obtain higher precision, and each flare was verified by a human by visual inspection to minimise false positive detections (e.g. accretion events, non-periodic variations) and ensure the purity of the catalogue.
We used all the stars with detected flares ($15\,124$), including the stars with real, but erroneously extracted flares (see Sect.\,2.5 of \citealt{2025A&A...694A.161S} for details).

We searched for radio counterparts in VLA Sky Survey \citep[VLASS,][]{2020RNAAS...4..175G,2021ApJS..255...30G} data, as at the time of the analyses, three separate epochs of observations were completed and available. The survey was conducted in three separate epochs between $2017$ and $2024$, with two observational runs in each, depending on the sky area, and is complete at declinations $\delta\ge-40\degr$. Detailed information on the observational runs is available at the survey website\footnote{\url{https://science.nrao.edu/vlass}}. The angular resolution of the survey is $2\farcs5$, and the  $5\sigma$ detection threshold can be as low as $\sim0.3$~mJy~beam$^{-1}$, depending on the local root-mean-square (rms) noise level of the radio maps. Out of all sources in samples BD and FS, $11\,519$ and $11\,154$ are in the footprint ($\delta\ge-40\degr$) of the VLASS, respectively. We searched radio images for these positions at all three epochs, and obtained single-epoch `quick look' radio maps from the Canadian Initiative for Radio Astronomy Data Analysis data base\footnote{\url{http://cutouts.cirada.ca/}} (CIRADA) using the \mbox{\code{astroquery}} package. We collected all available $25''\times25''$ radio map cutouts centred on the catalogue star positions (J2015.5 for BD, J2000 for FS). Using the proper motion values of the object and the exact  observational time from each of the corresponding VLASS images, we extrapolated the co-ordinates of the star at the time of the radio measurements. We searched for any radio emission associated with the target sources in the  $10\arcsec\times10\arcsec$ cutout images centred on these `updated' positions. We defined an emission feature as a radio detection, where the cutout image had a signal-to-noise ratio of SNR $\ge6$, considering the noise levels as the root mean square value of the corresponding radio map. As it was discussed by \citet{2002AJ....124.2364I}, for cross-matching Faint Images of the Radio Sky at Twenty-centimeters \citep[FIRST,][]{1997ApJ...475..479W,2015ApJ...801...26H} survey radio images (angular resolution of $5''$), separations above $2\farcs5$ significantly increase the number of random associations not corresponding to the optical counterpart, and a search radius of $1\farcs5$ was advised. Considering the finer angular resolution ($2\farcs5$) of the VLASS, the astrometric precisions of both \textit{Gaia} \cite[$\lesssim0.5$~mas,][]{2018A&A...616A...1G,2023A&A...674A...1G}  and the VLASS quick look images \citep[$0\farcs5$ at $\delta\ge-20\degr$, $1\arcsec$ at $\delta<20\degr$,][]{2021ApJS..255...30G}, and the angular sizes of the stellar diameters, which are within the sub-mas range, we would not expect to find associated radio emission beyond the VLASS positional uncertainties. Thus we defined the maximum separation of the optical and radio positions as $\sqrt{2}\arcsec$ for a match, corresponding to coincidence within $1\arcsec$ in both right ascension and declination.

For additional radio measurements, the list of the $55$ objects were cross-matched with the FIRST survey and the Sydney radio star catalogue \citep[D24,][]{2024PASA...41...84D}, constructed from radio-identifications in the Low Frequency Array \cite[LOFAR,][]{2013A&A...556A...2V} and the Australian SKA Pathfinder \citep[ASKAP,][]{2021PASA...38....9H} surveys. We applied search radii of $1\farcs5$ and $5''$, respectively, for the FIRST survey and D24 cross-match.

To cross-match stellar activity in the radio and the optical wavebands, $2$-minute cadence Science Processing Operations Center (SPOC) \textit{TESS} light curves were utilised, which were acquired using the \code{lightkurve} package. To estimate the optical flare rate of stars detected by VLASS, we identified flares on the \textit{TESS} light curves manually, using a box-selection tool to mark the flaring points on the light curve. As the sample includes young stars with erratic brightness variations, automated algorithms would struggle to find flares and exclude false positives. We calculated the energies of these flares in the \textit{TESS} band ($E_{\rm TESS}$) by integrating the area under the normalised light curve, and multiplying it with the quiescent luminosity of the star in the \textit{TESS} band, with the method described in Sect.~2.8 of \cite{2025A&A...694A.161S}. To summarise the optical flaring activity of each star in a simple way, we defined the flare rate as the number of observed flares per day. To make the flare rates of different stars comparable, we only counted the flares above the threshold of $E_{\rm TESS} > 10^{34}$\,erg. Flare rates defined as the fraction of time the star spends in flaring state were also determined (irrespective of $E_{\rm TESS}$). These values were not used in the analyses; however, they are provided in the extended version of Table~\ref{tab:radiox} as supplementary material. We checked the NASA Exoplanet 
Archive\footnote{\url{https://exoplanetarchive.ipac.caltech.edu/}}
to see whether a planetary companion could influence the activity properties of the hosts, but only TIC 268397995 (V830 Tau) has a known exoplanet from our targets -- unfortunately, a sample size of one does not allow us to perform a detailed statistical study. 

For comparison, radio stars non-identical with our final sample were used from D24. For simplicity, only catalogue data were used from the VLASS survey. As the exact observation dates are not included in the catalogues, proper motion corrected co-ordinates were calculated for the mid-dates of the first and second VLASS epoch, August 2018 and March 2021, respectively.  To ensure that the positions do not change drastically during the $\sim2$~yr long VLASS runs, thus rendering the cross-matching process unreliable, only stars with low proper motion ($\mu_\mathrm{RA}<200$~mas~yr$^{-1}$ and $\mu_\mathrm{DEC}<200$~mas~yr$^{-1}$) were used ($777$ stars). This way the uncertainty of the co-ordinates within the survey time range are maximised at $\sim0\farcs5$.

\section{Results}
\subsection{Radio counterparts in the VLA Sky Survey footprint}

\begin{figure}[t]
    \centering
    \includegraphics[width=0.9\linewidth]{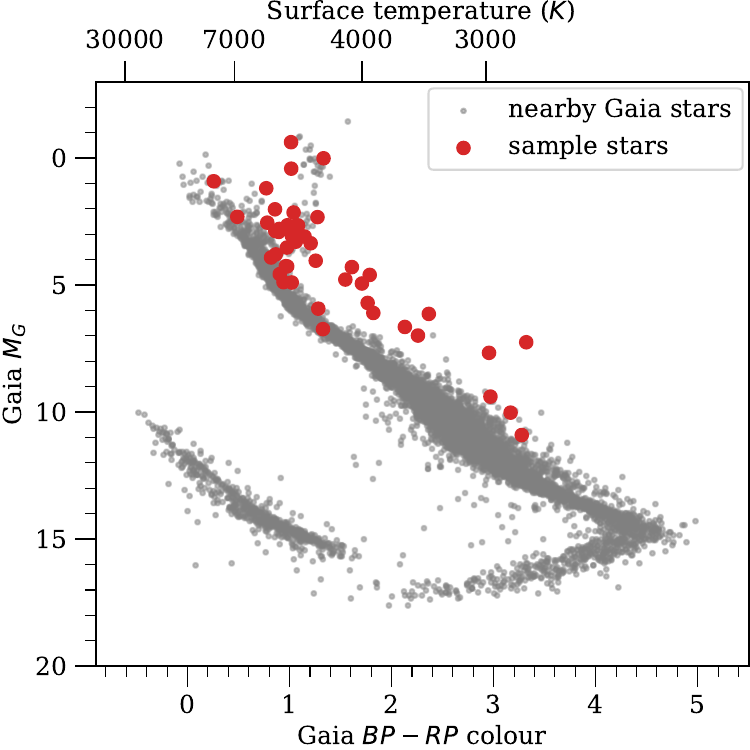}
    \caption{Colour--magnitude diagram of the $55$ radio-detected sample stars with nearby ($<50$~pc) \textit{Gaia} stars. The loci of the sample stars were corrected for interstellar extinction with the \code{seismolab} package \citep{Bodi2024}}.
    \label{fig:hrd}
\end{figure}

We did not find radio emission within the defined distance($\sqrt{2}\arcsec$) with respect to the proper motion-corrected co-ordinates of any of the sources in BD, while $55$ sources were found in FS to have close counterparts in either of the VLASS maps. The full list is provided in Table~\ref{tab:radiox} in Appendix~\ref{sec:radio_counterpart_app}. Figure~\ref{fig:hrd} shows these stars on the \textit{Gaia} colour--magnitude diagram. Many stars lie above the main sequence, suggesting that they are young objects. Based on the renormalised unit weight error (RUWE) parameter from \textit{Gaia} DR3, 19 objects are likely unresolved binaries (RUWE $>1.4$, see e.g. \citealt{2024A&A...688A...1C}). Radio flux densities were acquired from the VLASS radio component catalogues\footnote{Component catalogues for the first and second epoch VLASS observations are available at \url{https://cirada.ca/vlasscatalogueql0}}, versions v3 and v2 for the first and second epoch data, respectively. For the third epoch images, flux densities were obtained by fitting elliptical Gaussian model components to the images  using the \code{jmfit} task of the US National Radio Astronomy Observatory (NRAO) Astronomical Image Processing System \citep[\code{aips},][]{2003ASSL..285..109G} package. Upper limits on the flux density were considered as the $5\sigma$ rms noise level of the corresponding radio image for epochs where the source was not directly detected (SNR $<6$). Out of the $55$, four were found to have counterparts in the FIRST survey, and $18$ are also included in the full D24 sample. For the remaining stars from D24 (comparison sample),  $85$ and $93$ VLASS associations were found in the first and second epoch catalougues, respectively, out of which $54$ were detected at both epochs.

The level of radio variability was estimated by calculating the radio variability index, following the equation from \cite{1992ApJ...399...16A}:
\begin{equation}
    V=\frac{(S_\mathrm{max}-\sigma_{S_\mathrm{max}})-(S_\mathrm{min}+\sigma_{S_\mathrm{min}})}{(S_\mathrm{max}-\sigma_{S_\mathrm{max}})+(S_\mathrm{min}+\sigma_{S_\mathrm{min}})},
\end{equation} 
where $S_\mathrm{min}$ and $S_\mathrm{max}$ are the minimum and maximum flux density values, and $\sigma_{S_\mathrm{min}}$ and  $\sigma_{S_\mathrm{max}}$ are their respective errors. We found that $43$ out of the $55$ stars ($78\%$) have a variability index greater than $10\%$, and $16$ of them are highly variable in the radio wavebands ($V\ge0.5$). The D24 sample seems slightly less variable, with $31$ of the $54$ having variability indices $V>0.1$ and only three above $V>0.5$; however, this could be due to the inclusion of only the first two VLASS epochs.

Two stars from the $18$ shared with the D24 sample have more than one measurement in the same frequency band ($888$~MHz or $1.4$~GHz), TIC~241351537 and TIC~353053003. These two sources also were detected in the FIRST survey. The low-frequency variability for these sources were calculated. The $888$~MHz and $1.4$~GHz variability indices for TIC~241351537 were found $V_{888}=0.11$ and $V_{1.4}=0.01$, while for TIC~353053003 the values are $V_{888}=0.05$ and $V_{1.4}=0.18$, respectively. For comparison, the VLASS variability of these sources are $0.51$ and $0.06$.

\begin{figure*}
    \centering
    \includegraphics[width=\linewidth]{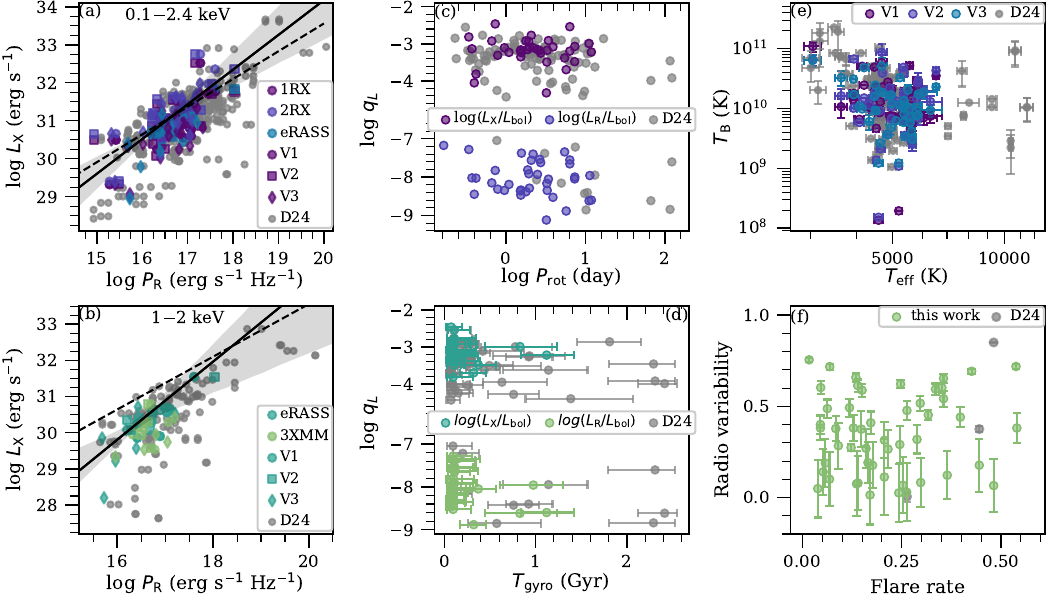}
    \caption{(a--b) Relation of the radio powers and X-ray luminosities final sample of flaring stars. Colour coding denotes the different X-ray survey data, while the three VLASS observational epochs are denoted with circle, square, and diamond markers. The solid black lines and shaded grey regions denote the best-fit model and the 95\% confidence interval (see Sect.~\ref{sec:rx_properties}), while the dashed lines denote the model by \citet{2014ApJ...785....9W} (a) The first and second \textit{ROSAT} survey data, and the \textit{eROSITA} are denoted with violet, blue, and teal, respectively. (b) Th \textit{eROSITA} and \textit{XMM-Newton} observations are denoted with dark and light green, respectively. (c) Relation of the period and the ratio of the radio/X-ray luminosities to the bolometric luminosity. (d) Relation of the estimated gyrochronological ages and the ratio of the radio/X-ray luminosities to the bolometric luminosity. (e) Relation between the effective and brightness temperatures. (f) Radio variability as the function of the flare rate. The comparison sample from \citet{2024PASA...41...84D} is shown with grey circles in each panel.}
    \label{fig:analytics}
\end{figure*}

\subsection{Radio and X-ray properties of flaring stars}\label{sec:rx_properties}

We found that all $55$ stars in the final sample of radio-emitting flare stars can also be associated with X-ray emission, in the point source catalogues of either the \textit{ROSAT} \citep{1999A&A...349..389V,2000IAUC.7432....3V,2016A&A...588A.103B} all-sky survey, the \textit{XMM-Newton} serendipitous survey \citep{2016A&A...590A...1R}, or the \textit{SRG/eROSITA} all-sky survey \citep{2024A&A...682A..34M}. Out of the $55$ X-ray detected stars, there is available distance information for all objects except for TIC~68416596, thus radio- and X-ray luminosities were calculated for the remaining $54$. We note that flux values from the second \textit{ROSAT} survey catalogue  $F_\mathrm{2RX}\gtrsim3\times10^{26}$~erg~s$^{-1}$~cm$^{-2}$ (ten times the maximum value of the first \textit{ROSAT} survey fluxes) were considered outliers, and no luminosity value was calculated for them. 
We found that they are in agreement with the radio-X-ray correlation, $L_\mathrm{X}/P_\mathrm{R}\sim10^{-15}$~[Hz$^{-1}$], proposed by e.g. \cite{1990SoPh..130..265B} and \cite{1993ApJ...405L..63G} (see Fig.~\ref{fig:analytics} panels a and b), suggesting that the detected $3$~GHz radio emission is most likely the quiescent component, indicating the level of the overall magnetic activity of the flaring stars. Fitting the logarithm of the radio powers and the $0.1-2.4$~keV  X-ray luminosities, we found that stars in our sample follow the relation of $\log L_\mathrm{X}=(0.89\pm 0.14) \log L_\mathrm{R} + (16.67 \pm 1.96)$. Excluding the six M dwarfs from the sample, the radio--X-ray relation is modified to $\log L_\mathrm{X}=(0.69 \pm 0.13 ) \log L_\mathrm{R} + ( 19.65 \pm 2.12)$. Repeating the analysis with the $1-2$~keV X-ray data points, the relation follows a slightly less steep slope of $\log L_\mathrm{X}=(0.99\pm 0.16) \log L_\mathrm{R} + (14.01\pm 2.68)$. Our results agree with those parameters found by \citep{2014ApJ...785....9W} and \cite{2024AJ....168..288A} within errors. We note that out of the $55$ stars in our analysis, there are $36$ objects  (within a search radius of $5''$) coinciding with the sample of \cite{2024AJ....168..288A}. On the contrary, many stars from the D24 comparison sample fall below the G{\"u}del--Benz relation, which can be contributed to the coherent nature of radio emission \cite[e.g.][]{2021ApJ...919L..10P,2022ApJ...926L..30V,2024PASA...41...84D}. This  provides further evidence for the incoherent synchrotron origin of the radio emission of our flaring star sample.

X-ray and radio activity indices were calculated as $q_{L_\mathrm{X}}=L_\mathrm{X}/L_\mathrm{bol}$ and $q_{L_\mathrm{R}}=L_\mathrm{R}/L_\mathrm{bol}$, respectively. Activity indices as the function of the rotational period are shown in Fig.~\ref{fig:analytics} panel c. Rotational periods were calculated utilising \textit{TESS} light curves and the \code{lightkurve} package. Bolometric luminosities were calculated with $V$-band magnitudes from \citet{2019AJ....158..138S}, while the bolometric corrections were applied following the relation from \citet{1996ApJ...469..355F}. The X-ray activity of the stars in our sample follows the activity--rotation period relation described by e.g. \citet{2003A&A...397..147P,2019A&A...628A..41P}, and a similar trend was found for the radio activity for the sample. X-ray emission saturation occurs below the critical rotational period value $P_\mathrm{rot}\sim10$~days, where an X-ray luminosities are limited to a a maximum level (saturation) of $\log(L_\mathrm{X}/L_\mathrm{bol})\approx-3$ \citep{2003A&A...397..147P,2011ApJ...743...48W}. 
As apparent in Fig.~\ref{fig:analytics} panel c, all of the flaring stars in our final sample are in the saturation regime. We note that the time base of the \textit{TESS} light curves is only sensitive to short period variabilities, and is not suitable to identify long-term effects. As the radio activity index shows a similar behaviour, a common mechanism for the origin of the X-ray and radio emissions is further supported. The D24 sample shows similar activity indices as our flaring stars, only three of them being in the non-saturated regime, with $\sim100$~day rotational periods.

Radio brightness temperatures were calculated following the equation from \citet{2002ARA&A..40..217G}. We show the brightness temperatures as a function of the effective temperatures in Fig.~\ref{fig:analytics} panel e. Although there is a slight hint of a anti-correlation of the two parameters, no quantitative correlation was found ($r<0.2$). Including the temperature of the D24 stars, the possibility of a connection is more apparent, $\log T_\mathrm{b}=(-1.3\pm0.5)\cdot \log T_\mathrm{eff}+(15.0\pm1.8)$ ($r=-0.3$,$p=0.007$). All of the brightness temperature values in the sample exceed the thermal--synchrotron limit of $10^4$~K \citep[e.g.][]{1982ApJ...252..102C}, are in agreement with the values found for other flaring stars \citep[e.g.][]{2015MNRAS.446.3687P}, and are in the incoherent emission region \citep[$T_\mathrm{b}<10^{12}$~K, e.g.][]{2024AJ....168..288A}. Based on the findings above, we can conclude that the radio emission has a synchrotron origin, and is most likely connected to the flaring activity itself, or to the heating processes of the stellar corona.

\subsection{Evolutionary state of the sample}
Using the \textsc{gyrointerp} package \citep{2023ApJ...947L...3B}, gyrochronological ages were estimated for $31$ out of the $55$ flaring stars, all with known effective temperatures within the range of $3800-6200$~K. We found that all $31$ flaring stars are at a relatively early evolutionary stage ($T_\mathrm{gyro}\lesssim1$~Gyr, $T_\mathrm{gyro,median}=80$~Myr, standard deviance $\sigma\sim250$~Myr), with the majority ($26$; $84\%$) of the sources being younger than $150$~Myr (Fig.~\ref{fig:analytics}, panel d).  We found that both the X-ray and radio activity indices show a similar `non-dependence' on the gyrochronological age estimates, as was seen with the activity--rotation relation (Fig.~\ref{fig:analytics}, panel c). This implies that flaring stars with sufficient activity in both the radio and X-ray regimes are the members of the youngest stellar population.
The median age estimates of the D24 stars are similar ($90$~Myr); however the age distribution has a higher standard deviation ($\sigma\sim500$~Myr). D24 stars are slightly older than our flaring stars, as only $63\%$ of them are younger than $150$~Myr, with the oldest ones surpassing the age of $2$~Gyr. We note that the large spread found in the rotation rates for the youngest stellar population \citep{2015A&A...577A..28J} raises caution for interpreting the estimated values of stellar ages. Similarly, subgiant stars may also deviate from traditional rotation--age relations \citep{2014A&A...572A..34G}, making the age estimates less reliable in these cases.

Stars evolve through multiple distinct phases that influence their observable behaviour, and this evolution is best visualised on the Hertzsprung–Russell diagram (HRD). Formation begins in molecular clouds, where dense clumps collapse into protostars. These evolve through the pre-main sequence (PMS) phase along the Hayashi track, becoming hotter and more compact as a radiative core develops. Once hydrogen fusion begins in the core, the star reaches the zero-age main sequence (ZAMS). On the main sequence (MS), high-mass and low-mass stars differ in internal structure and fusion processes (CNO cycle vs. pp-chain). When core hydrogen is depleted, fusion continues in a shell, and the star ascends to the sub-giant and then red giant branch (RGB), developing a convective envelope and expanding significantly. In low-mass stars, helium accumulates in a degenerate core, leading to an off-centre helium flash, causing a thermonuclear runaway and energy loss. Repeated flashes reduce core degeneracy, culminating in stable helium burning and hydrogen shell fusion on the horizontal branch, marking the end of the RGB phase.

Our flaring stars with associated radio emission can be found at three evolutionary phases on the HRD (Fig.~\ref{fig:hrd}). The majority of them are either young stellar objects (YSOs) at the end of the Hayashi track, or are subgiants at their turn-off points just leaving the MS. We can also find a small number of F--G spectral type MS stars in the sample. The presence of a large number of YSOs in our sample is also supported by the estimated gyrochronology ages.

\subsection{Relationship of radio emission and stellar parameters}

\begin{figure}
    \centering
    \includegraphics[width=0.65\linewidth]{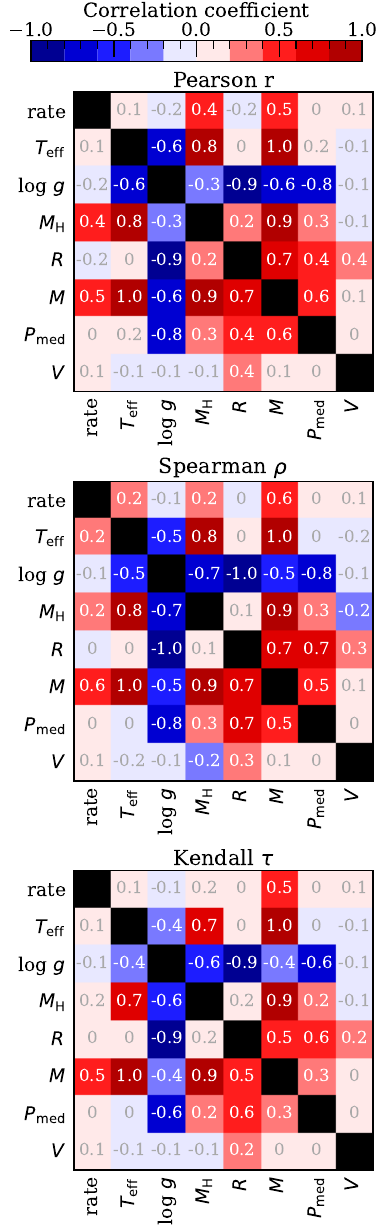}
    \caption{Correlation matrices showing the relation between the stellar parameters (flare rate, effective temperature, surface gravity, metallicity, radius, mass) and radio properties (median value of the radio powers, radio variability index) of the final list of flaring stars.}
    \label{fig:correlation}
\end{figure}

To examine the connection between the radio emission and other stellar parameters, we performed correlation analysis, using the following quantities: flare rate, effective temperature, surface gravity, metallicity, radius, mass, median value of the radio powers, and radio variability index. Stellar parameters were acquired from the \textit{TESS} input catalogue \citep{2019AJ....158..138S} and the \textit{Gaia} DR3 catalogue \citep{2016A&A...595A...1G,2023A&A...674A...1G,2023A&A...674A..32B}. The correlation ranks between the stellar parameters and the radio emission were examined using the Pearson-r, Spearman-$\rho$, and Kendall-$\tau$ correlation coefficients (Fig.~\ref{fig:correlation}).
The radio powers show an anti-correlation with the logarithm of the surface gravity, and a positive correlation with the stellar radius and mass, following the linear trends
\begin{gather}
    \log P_\mathrm{med}=(-0.87\pm0.17)\cdot \log g + (12.70\pm0.71),\\
    \log P_\mathrm{med}=(1.04 \pm 0.17) \cdot \log R+ (9.06 \pm 0.08),\\
    \mathrm{and}~\log P_\mathrm{med}=(1.20\pm0.39)\cdot M + (8.05\pm0.39).
\end{gather}

The connection between the median radio powers and the stellar parameters ($\log g$, $R$, $M$) is supported by all three of the Pearson, Spearman, and Kendall correlation coefficients, with a significance value of $p < 0.05$ in all cases. 
We note that there is an apparent correlation between the radio powers and the metallicity as well; however, all three tests have $p>0.3$, thus the relation cannot be confirmed. There is no strong correlation between the radio variability and the flaring rate in the sample (see also Fig.~\ref{fig:analytics}, panel f), and only a weak correlation was found between the flare rate, effective temperature, metallicity, and mass; however, neither of them is statistically significant.

\subsection{Relationship of flaring activity and radio emission for individual sources}

\begin{figure*}[ht]
    \centering
    \includegraphics[width=0.75\linewidth]{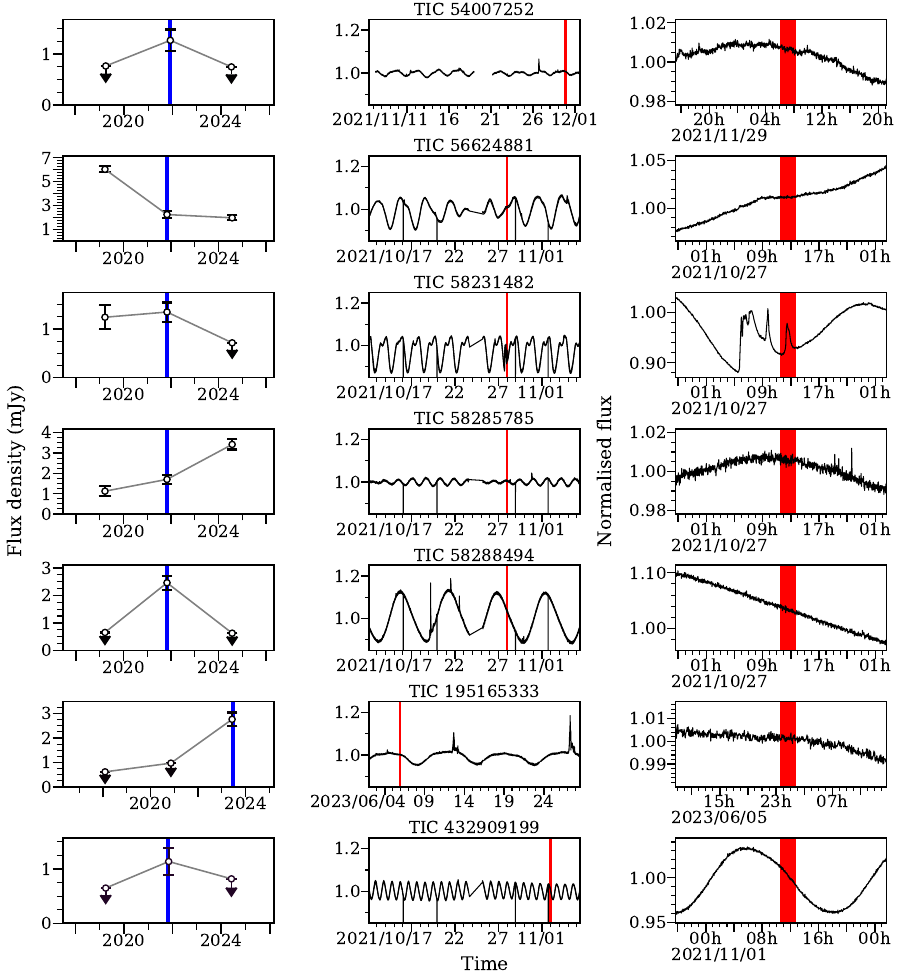}
    \caption{Radio detected flaring stars with coincident \textit{TESS} observations. \textit{Left:} Radio flux density values during the three VLASS observation epochs.  Non-detections are denoted as upper limits at $5\sigma$ rms noise levels. Blue lines denote the time span of the \textit{TESS} observations coincident with the VLASS scans. \textit{Middle:} \textit{TESS} light curves. The red lines denote time span the VLASS measurements coincident with \textit{TESS} observations. \textit{Right:} 16-h-zoom-ins of the \textit{TESS} light curves. The red areas denote the $\sim2$-h time span of the VLASS measurements.}\label{fig:lightcurves7}
\end{figure*}

Out of the $55$ flaring stars, seven were found to have \textit{TESS} light curves coincident with VLASS observations of the same object. These sources are listed in Table~\ref{tab:tess_radio7}, and their radio- and optical light curves are shown in Figure~\ref{fig:lightcurves7}. Although flaring activity can be identified in all of these sources in the $\sim1$~month long \textit{TESS} light curves, only one of them was  detected during the $\sim2$~h scan of the corresponding VLASS sky region, TIC~58231482. None of these seven flaring stars were covered by the FIRST survey footprint, and three sources were found to have radio counterparts in D24, TIC~56624881, TIC~58285785, and TIC~195165332. Although TIC~56624881 and TIC~58285785 have available data at multiple wavebands, unfortunately, neither of them have multi-epoch observations.

\begin{table}[th] \setlength{\tabcolsep}{4pt}
\caption{Radio detected flaring stars with coincident \textit{TESS }observations. }\label{tab:tess_radio7}
    \begin{tabular}{llrrrc}
\hline\hline
ID	&	Rate	&	$S_{1}$~(mJy)			&	$S_{2}$~(mJy)			&		$S_{3}$~(mJy)		&	$V$	\\ \hline
TIC~54007252	&	0.041	&	<0.7	  		&	1.3	$\pm$	0.2	&	<0.8	  		&	0.08	\\
TIC~56624881	&	0.042	&	6.0	$\pm$	0.2	&	2.2	$\pm$	0.3	&	2.0	$\pm$	0.2	&	0.45	\\
TIC~58231482	&	0.052	&	1.2	$\pm$	0.3	&	1.3	$\pm$	0.2	&	<0.7	  		&	0.12	\\
TIC~58285785	&	0.041	&	1.1	$\pm$	0.2	&	1.7	$\pm$	0.2	&	3.4	$\pm$	0.3	&	0.40	\\
TIC~58288494	&	0.034	&	<0.6	  		&	2.5	$\pm$	0.2	&	<0.6	  		&	0.49	\\
TIC~195165333	&	0.072	&	<0.6	  		&	<0.9	  		&	2.8	$\pm$	0.3	&	0.54	\\
TIC~432909199	&	0.018	&	<0.6	  		&	1.1	$\pm$	0.2	&	<0.8	  		&	0.07	\\
\hline
    \end{tabular}   
{\\\\ \textit{Notes:} Column 1 -- \textit{TESS} Input catalogue number of the star, Column 2 -- flare rate, Column 3--5 -- flux density at 3~GHz in the VLASS 1--3 epoch observations. Upper limits for the non-detections are given at $5\sigma$ rms noise levels, Column 6 -- radio variability index.}
\end{table}

\begin{description}[leftmargin=0cm,itemsep=5pt]
\item[TIC~54007252 (HD~242903, V1362 Tau)]
is an RS~CVn type variable \citep{2003AstL...29..468S} with no published radio detection to date. This could be due its relatively low flux density, and a higher radio variability than estimated from the upper limit on the non-detection at the first and third VLASS epochs. As there is more than a day delay between the detected optical flare and the VLASS radio observations, a connection between the two is not expected. As the radio emission of RS~CVn stars is generally known to be highly variable \citep[e.g.][and references therein]{2003A&A...403..613G}, and it is believed to originate from magnetic field related gyrosynchrotron processes \citep[e.g.][]{1985ARA&A..23..169D,1994SSRv...68..293L}, flux density variability is not expected to correlate with the optical flaring episodes.

\item[TIC~56624881 (HD 283447, V773 Tau)] 
was classified as an Orion-type variable star \citep{2012ApJ...756...27L} and was found to be a quadrupole system \citep[e.g.][]{2012ApJ...747...18T}. TIC~56624881 was detected at all three VLASS epochs, and has additional radio flux density measurements at four frequency bands between $888$~MHz and  $1.7$~GHz observed by ASKAP \citep{2024PASA...41...84D}. It was also investigated at mas scales with the Very Long Baseline Array \citep[VLBA,][]{2012ApJ...747...18T,2018ApJ...859...33G,2023A&A...676A..11L}. The apparent separation between subsystems A and B is $\sim 150$~mas \cite{1993AJ....106.2005G,1993A&A...278..129L}. Component A was found to be a strong radio source and a spectroscopic binary itself \citep[e.g.][]{2007ApJ...670.1214B,2008A&A...480..489M}, with both Aa and Ab subcomponents detected at mas scales \citep{2018ApJ...859...33G,2023A&A...676A..11L}. Component Aa is highly variable, with a variability index of $0.61$ at 5~GHz and $0.96$ at 8~GHz, while component Ab shows a slightly less variable 5-GHz radio emission with $V=0.39$. 

\item[TIC~58231482 (HD 283518, V410 Tau)]
is a multiple system \citep{2012ApJ...751..115H}, known as a radio star \citep[e.g][]{1982ApJ...253..707C,1984ApJ...282..699B,2004A&A...427..263F}, with only one flare detected in the radio waveband \citep{1982ApJ...253..707C,2004A&A...427..263F}. The star was also classified as an Orion variable \citep{2012ApJ...756...27L}. Only one of the multiple components was detected with very long baseline interferometry (VLBI) with flux density varying between $1.21\pm0.04$ and $3.59\pm0.06$ at $5$~GHz, and $0.72\pm0.09$ and $13.62\pm0.22$ at $8$~GHz \citep{2018ApJ...859...33G}. 
Out of the seven radio-matched stars with simultaneous \textit{TESS} measurements, this is the only one with an optical flare captured within the respective $2$-h VLASS observing window. This optical flare is complex event consisting of five peaks. It is not affected by \textit{TESS} systematics, the flare is present on the light curves of four different data providers (120\,s and 600\,s cadence SPOC, 600\,s QLP, 1800\,s CDIPS). The last peak of the complex flare coincided with the second epoch VLASS scan; however, TIC~58231482 shows no significant brightening (Fig.~\ref{fig:lightcurves7}) when compared to the first epoch value. This implies no clear correlation between the radio variability and the optical flaring activity. On the other hand, as there is a $\sim50\%$ flux density decrease after the second VLASS epoch, a flaring activity captured at both the first and second epochs could explain the absence of increase in the brightness of the star. 
The mas-scale variability indices based on the $5$ and $8$-GHz VLBA monitoring observations in $2012$ \citep{2018ApJ...859...33G} are $V=0.48$ and $V=0.89$, respectively. These values are considerably higher than the $3$-GHz VLASS value of $V=0.12$. The high variability in the 2012 data can also be interpreted as the effect of enhanced flaring activity in the radio regime. 

\item[TIC~58285785 (V1023 Tau)]
is an Orion variable \citep{2012ApJ...756...27L} and
a spectroscopic binary \citep{2012ApJ...745..119N}.  Both stellar components were detected with the VLBA in the $2012$ monitoring observations \citep{2018ApJ...859...33G}. Component A shows clear variability, with variability indices of  $V=0.34$ and $V=0.92$ at $5$ and $8$~GHz, respectively, while component B was only detected at two observing epochs only at $5$~GHz, implying no change in the radio flux densities \citep{2018ApJ...859...33G}. It was detected at all three VLASS epochs, and has low-frequency detections with ASKAP at 3 frequency bands between $944$~MHz and $1.7$~GHz \cite{2024PASA...41...84D}. We found that the VLASS variability of the system is $V=0.40$,  and that the radio light curve shows $300\%$ flux density brightening through the three VLASS epochs (Fig.~\ref{fig:lightcurves7}), which implies an increase in the stellar activity. Based on the correlation described in e.g. \citet{2014MNRAS.441.2744V}, the $1.56$~days rotational period of the star would translate to an estimated minimum span of $\sim1-2$~years for the stellar activity cycle, agreeing with the timescale of the VLASS brightening; however, the time base of the observations is not enough to confirm this suggestion.

\item[TIC~58288494 (V819 Tau)]
is a weak line T Tauri type variable \citep{2008AJ....135..850D,2015MNRAS.453.3706D}. It has a radio variability index of $V=0.49$; however, as it was detected only at the second VLASS epoch in $2021$, the real value could be even greater. The star was targeted with VLBI observations in $2012$, but was not detected at either $5$ or $8$~GHz, with an upper limit on the radio intensity of $I<0.13$~mJy~beam$^{-1}$ \citep{2018ApJ...859...33G}. The non-detection at both arcsec and mas scales in the majority of the observing epochs could also support the idea that the detection at $2021$ is a radio outburst connected to the flaring activity; however, it does not coincide with any flares in the optical domain (Fig.~\ref{fig:lightcurves7}).  As the young star is still in its protostellar disc phase of evolution \citep{2021ApJ...920..132P}, the extreme radio changes can also be attributed to episodes of intermittent accretion onto the young stellar object \citep[e.g.][]{2020ApJ...904...37W,2025ApJ...981...68R}.

\item[TIC~195165333 (HD 321269)] 
is a rotationally variable spotted spectroscopic binary (RS CVn) \citep{2012AcA....62...67K, 2022yCat.1358....0G,2023A&A...674A...1G}. It was detected only at the third VLASS epoch in $2023$. Considering the upper limits on the flux densities, this translates to a brightening of at least $150\%$ (Fig.~\ref{fig:lightcurves7}, Table~\ref{tab:tess_radio7}). The star was also detected earlier that year with ASKAP at $888$~MHz with a flux density of $4.2\pm0.9$mJy \citep{2024PASA...41...84D}. There are no optical flares coinciding directly with the $2$~h span of the VLASS observation window; however, two optical flares are captured by \textit{TESS} within a $24$-h time frame prior to the radio detection. This implies that the increase in the radio emission might be linked to the optical flaring activity.

\item[TIC~432909199 (HD 244354, V1368 Tau)]
is a class III (disc free) young stellar object \citep{2017ApJ...838..150K,2021ApJS..254...20L}. It was only detected at the second VLASS epoch. Considering the upper limits on the flux densities at non-detections, this would translate to a brightening of at least $170\%$, after which the star faded beyond the detection limit (Fig.~\ref{fig:lightcurves7}). We found no evidence for any optical flaring within a reasonable timescale close to the radio detection. The radio variability index is $V=0.07$; however, the real value might be considerably higher, as the non-detections could be attributed to the sensitivity threshold of the VLASS survey. Despite the lack of radio--optical connection, the brightening in epoch two can be attributed to flaring events captured in the radio regime without an optical counterpart.
\end{description}

\section{Summary and conclusions}

We collected two large samples to test the stellar activity in the radio regime, searching for radio counterparts within $\sqrt{2}''$ separation in the VLA Sky Survey data. We found no associated radio emission to any of the brown dwarf candidates from \citet{2018A&A...619L...8R}, and $55$ matches in the list of intermediate-to-late spectral type flaring stars from \citet{2025A&A...694A.161S}, out of which seven stars with coincident short-cadence \textit{TESS} light curves available were individually inspected in detail. 

Our study indicates that stars that are detectable in radio fall into three categories: YSOs at the end of the Hayashi track -- this is also suggested by the gyrochronology age estimation; subgiants leaving the MS; and a somewhat smaller number of F--G type MS stars. The YSOs are probably detectable because of their higher activity level, while the subgiant and MS samples are likely visible in radio due to their larger intrinsic luminosity (compared to cooler stars).  

We investigated the relationship of the radio activity with respect to the X-ray emission, flaring activity, rotational period, and age of the $55$ radio-emitting dwarf stars. Our statistical analysis revealed a strong relation between the radio powers and stellar parameters, such as the surface gravity, radius, and mass. We found no correlation between the flare rate and the radio variability index of the sample stars. The sample follows the radio--X-ray luminosity relation, suggesting a common underlying process for the two types of emission, and implying that the origin of the radio emission is of gyrosynchrotron or cyclotron nature. The final sample showed that radio-detected flaring stars belong to a predominantly young stellar population ($T_\mathrm{gyro}\lesssim1$~Gyr), with the majority having gyrochronological age estimates $T_\mathrm{gyro}<150$~Myr. We compared our sample to the LOFAR--ASKAP radio stars from \citet[][D24]{2024PASA...41...84D}, and found that the properties of the two sets of stars are consistent, with the exception of some D24 stars not following the G{\"u}del--Benz relation. Based on the individually examined sample of seven stars, we found that the individual flares do not show strict correlation with the changes in the radio regime; however, the connection cannot be ruled out entirely. We conclude that the radio emission detected for intermediate-to-late type flaring dwarf stars originate from multiple processes, including stellar flares, coronal heating, and possibly accretion onto a young star.  

\section*{Data availability}
An extended version of the final list of flaring stars (Table~\ref{tab:radiox}) is also provided as supplementary material, and is available at the CDS via anonymous ftp to cdsarc.cds.unistra.fr (130.79.128.5) or via https://cdsarc.cds.unistra.fr/viz-bin/cat/J/A+A/.
 
\begin{acknowledgements}
We thank the anonymous referee for their constructive suggestions to improve our manuscript.
We acknowledge the support of the Hungarian National Research, Development and Innovation Office (NKFIH) Élvonal grant KKP 143986.
On behalf of the \textit{"Looking for stellar CMEs on different wavelengths"} project we are grateful for the possibility of using HUN-REN Cloud \cite{MTACloud} which helped us achieve the results published in this paper.

The National Radio Astronomy Observatory is a facility of the National Science Foundation operated under cooperative agreement by Associated Universities, Inc. CIRADA is funded by a grant from the Canada Foundation for Innovation 2017 Innovation Fund (Project 35999), as well as by the Provinces of Ontario, British Columbia, Alberta, Manitoba and Quebec. This research has made use of the NASA/IPAC Infrared Science Archive, which is funded by the National Aeronautics and Space Administration and operated by the California Institute of Technology. 

This work has made use of data from the European Space Agency (ESA) mission
{\it Gaia} (\url{https://www.cosmos.esa.int/gaia}), processed by the {\it Gaia}
Data Processing and Analysis Consortium (DPAC,
\url{https://www.cosmos.esa.int/web/gaia/dpac/consortium}). Funding for the DPAC
has been provided by national institutions, in particular the institutions
participating in the {\it Gaia} Multilateral Agreement.

This paper includes data collected with the \textit{TESS} mission. Funding for the \textit{TESS} mission is provided by the NASA Explorer Program. STScI is operated by the Association of Universities for Research in Astronomy, Inc., under NASA contract NAS 5–26555.

This work is based on data from eROSITA, the soft X-ray instrument aboard SRG, a joint Russian-German science mission supported by the Russian Space Agency (Roskosmos), in the interests of the Russian Academy of Sciences represented by its Space Research Institute (IKI), and the Deutsches Zentrum für Luft- und Raumfahrt (DLR). The SRG spacecraft was built by Lavochkin Association (NPOL) and its subcontractors, and is operated by NPOL with support from the Max Planck Institute for Extraterrestrial Physics (MPE). The development and construction of the eROSITA X-ray instrument was led by MPE, with contributions from the Dr. Karl Remeis Observatory Bamberg \& ECAP (FAU Erlangen-Nuernberg), the University of Hamburg Observatory, the Leibniz Institute for Astrophysics Potsdam (AIP), and the Institute for Astronomy and Astrophysics of the University of Tübingen, with the support of DLR and the Max Planck Society. The Argelander Institute for Astronomy of the University of Bonn and the Ludwig Maximilians Universität Munich also participated in the science preparation for eROSITA. 

This research has made use of data obtained from the 3XMM XMM-Newton serendipitous source catalogue compiled by the 10 institutes of the XMM-Newton Survey Science Centre selected by ESA. 

This research has made use of the VizieR catalogue access tool, CDS,
Strasbourg, France \citep{10.26093/cds/vizier}. The original description of the VizieR service was published in \citet{vizier2000}.

This research has made use of the NASA Exoplanet Archive, which is operated by the California Institute of Technology, under contract with the National Aeronautics and Space Administration under the Exoplanet Exploration Program.

This research has made use of the Astrophysics Data System, funded by NASA under Cooperative Agreement 80NSSC21M00561.

\end{acknowledgements}

\bibliographystyle{aa}
\bibliography{aa55032-25.bib}

\begin{thebibliography}{127}
\expandafter\ifx\csname natexlab\endcsname\relax\def\natexlab#1{#1}\fi

\bibitem[{{Ainsworth} {et~al.}(2014){Ainsworth}, {Scaife}, {Ray}, {Taylor}, {Green}, \& {Buckle}}]{2014ApJ...792L..18A}
{Ainsworth}, R.~E., {Scaife}, A. M.~M., {Ray}, T.~P., {et~al.} 2014, \apjl, 792, L18

\bibitem[{{Aller} {et~al.}(1992){Aller}, {Aller}, \& {Hughes}}]{1992ApJ...399...16A}
{Aller}, M.~F., {Aller}, H.~D., \& {Hughes}, P.~A. 1992, \apj, 399, 16

\bibitem[{{Alonso-Floriano} {et~al.}(2015){Alonso-Floriano}, {Morales}, {Caballero}, {Montes}, {Klutsch}, {Mundt}, {Cort{\'e}s-Contreras}, {Ribas}, {Reiners}, {Amado}, {Quirrenbach}, \& {Jeffers}}]{2015A&A...577A.128A}
{Alonso-Floriano}, F.~J., {Morales}, J.~C., {Caballero}, J.~A., {et~al.} 2015, \aap, 577, A128

\bibitem[{{Anglada} {et~al.}(2018){Anglada}, {Rodr{\'\i}guez}, \& {Carrasco-Gonz{\'a}lez}}]{2018A&ARv..26....3A}
{Anglada}, G., {Rodr{\'\i}guez}, L.~F., \& {Carrasco-Gonz{\'a}lez}, C. 2018, \aapr, 26, 3

\bibitem[{{Aronow} {et~al.}(2018){Aronow}, {Herbst}, {Hughes}, {Wilner}, \& {Winn}}]{2018AJ....155...47A}
{Aronow}, R.~A., {Herbst}, W., {Hughes}, A.~M., {Wilner}, D.~J., \& {Winn}, J.~N. 2018, \aj, 155, 47

\bibitem[{{Ayanabha} {et~al.}(2024){Ayanabha}, {Narang}, {Puravankara}, {Shridharan}, {Tyagi}, {Banerjee}, {Nayak}, \& {Surya}}]{2024AJ....168..288A}
{Ayanabha}, D., {Narang}, M., {Puravankara}, M., {et~al.} 2024, \aj, 168, 288

\bibitem[{{Babusiaux} {et~al.}(2023){Babusiaux}, {Fabricius}, {Khanna}, {Muraveva}, {Reyl{\'e}}, {Spoto}, {Vallenari}, {Luri}, {Arenou}, {{\'A}lvarez}, {Anders}, {Antoja}, {Balbinot}, {Barache}, {Bauchet}, {Bossini}, {Busonero}, {Cantat-Gaudin}, {Carrasco}, {Dafonte}, {Diakit{\'e}}, {Figueras}, {Garcia-Gutierrez}, {Garofalo}, {Helmi}, {Jim{\'e}nez-Arranz}, {Jordi}, {Kervella}, {Kostrzewa-Rutkowska}, {Leclerc}, {Licata}, {Manteiga}, {Masip}, {Mongui{\'o}}, {Ramos}, {Robichon}, {Robin}, {Romero-G{\'o}mez}, {S{\'a}ez}, {Santove{\~n}a}, {Spina}, {Torralba Elipe}, \& {Weiler}}]{2023A&A...674A..32B}
{Babusiaux}, C., {Fabricius}, C., {Khanna}, S., {et~al.} 2023, \aap, 674, A32

\bibitem[{{Bastian}(1990)}]{1990SoPh..130..265B}
{Bastian}, T.~S. 1990, \solphys, 130, 265

\bibitem[{{Benz} \& {G{\"u}del}(1994)}]{1994A&A...285..621B}
{Benz}, A.~O. \& {G{\"u}del}, M. 1994, \aap, 285, 621

\bibitem[{{Biazzo} {et~al.}(2012){Biazzo}, {Alcal{\'a}}, {Covino}, {Sterzik}, {Guillout}, {Chavarr{\'\i}a-K.}, {Frasca}, \& {Raddi}}]{2012A&A...542A.115B}
{Biazzo}, K., {Alcal{\'a}}, J.~M., {Covino}, E., {et~al.} 2012, \aap, 542, A115

\bibitem[{{Bieging} {et~al.}(1984){Bieging}, {Cohen}, \& {Schwartz}}]{1984ApJ...282..699B}
{Bieging}, J.~H., {Cohen}, M., \& {Schwartz}, P.~R. 1984, \apj, 282, 699

\bibitem[{{Boden} {et~al.}(2007){Boden}, {Torres}, {Sargent}, {Akeson}, {Carpenter}, {Boboltz}, {Massi}, {Ghez}, {Latham}, {Johnston}, {Menten}, \& {Ros}}]{2007ApJ...670.1214B}
{Boden}, A.~F., {Torres}, G., {Sargent}, A.~I., {et~al.} 2007, \apj, 670, 1214

\bibitem[{{Boller} {et~al.}(2016){Boller}, {Freyberg}, {Tr{\"u}mper}, {Haberl}, {Voges}, \& {Nandra}}]{2016A&A...588A.103B}
{Boller}, T., {Freyberg}, M.~J., {Tr{\"u}mper}, J., {et~al.} 2016, \aap, 588, A103

\bibitem[{{Bouma} {et~al.}(2023){Bouma}, {Palumbo}, \& {Hillenbrand}}]{2023ApJ...947L...3B}
{Bouma}, L.~G., {Palumbo}, E.~K., \& {Hillenbrand}, L.~A. 2023, \apjl, 947, L3

\bibitem[{{Bourg{\'e}s} {et~al.}(2014){Bourg{\'e}s}, {Lafrasse}, {Mella}, {Chesneau}, {Bouquin}, {Duvert}, {Chelli}, \& {Delfosse}}]{2014ASPC..485..223B}
{Bourg{\'e}s}, L., {Lafrasse}, S., {Mella}, G., {et~al.} 2014, in Astronomical Society of the Pacific Conference Series, Vol. 485, Astronomical Data Analysis Software and Systems XXIII, ed. N.~{Manset} \& P.~{Forshay}, 223

\bibitem[{{Bowler} {et~al.}(2019){Bowler}, {Hinkley}, {Ziegler}, {Baranec}, {Gizis}, {Law}, {Liu}, {Shah}, {Shkolnik}, {Riaz}, \& {Riddle}}]{2019ApJ...877...60B}
{Bowler}, B.~P., {Hinkley}, S., {Ziegler}, C., {et~al.} 2019, \apj, 877, 60

\bibitem[{Bódi(2024)}]{Bodi2024}
Bódi, A. 2024, Journal of Open Source Software, 9, 7118

\bibitem[{{Callingham} {et~al.}(2021){Callingham}, {Vedantham}, {Shimwell}, {Pope}, {Davis}, {Best}, {Hardcastle}, {R{\"o}ttgering}, {Sabater}, {Tasse}, {van Weeren}, {Williams}, {Zarka}, {de Gasperin}, \& {Drabent}}]{2021NatAs...5.1233C}
{Callingham}, J.~R., {Vedantham}, H.~K., {Shimwell}, T.~W., {et~al.} 2021, Nature Astronomy, 5, 1233

\bibitem[{{Cannon} \& {Pickering}(1993)}]{1993yCat.3135....0C}
{Cannon}, A.~J. \& {Pickering}, E.~C. 1993, {VizieR Online Data Catalog: Henry Draper Catalogue and Extension (Cannon+ 1918-1924; ADC 1989)}, VizieR On-line Data Catalog: III/135A. Originally published in: Harv. Ann. 91-100 (1918-1924)

\bibitem[{{Castro-Ginard} {et~al.}(2024){Castro-Ginard}, {Penoyre}, {Casey}, {Brown}, {Belokurov}, {Cantat-Gaudin}, {Drimmel}, {Fouesneau}, {Khanna}, {Kurbatov}, {Price-Whelan}, {Rix}, \& {Smart}}]{2024A&A...688A...1C}
{Castro-Ginard}, A., {Penoyre}, Z., {Casey}, A.~R., {et~al.} 2024, \aap, 688, A1

\bibitem[{{Cohen} {et~al.}(1982){Cohen}, {Bieging}, \& {Schwartz}}]{1982ApJ...253..707C}
{Cohen}, M., {Bieging}, J.~H., \& {Schwartz}, P.~R. 1982, \apj, 253, 707

\bibitem[{{Condon} {et~al.}(1982){Condon}, {Condon}, {Gisler}, \& {Puschell}}]{1982ApJ...252..102C}
{Condon}, J.~J., {Condon}, M.~A., {Gisler}, G., \& {Puschell}, J.~J. 1982, \apj, 252, 102

\bibitem[{{Davis} {et~al.}(2021){Davis}, {Vedantham}, {Callingham}, {Shimwell}, {Vidotto}, {Zarka}, {Ray}, \& {Drabent}}]{2021A&A...650L..20D}
{Davis}, I., {Vedantham}, H.~K., {Callingham}, J.~R., {et~al.} 2021, \aap, 650, L20

\bibitem[{{Devor} {et~al.}(2008){Devor}, {Charbonneau}, {O'Donovan}, {Mandushev}, \& {Torres}}]{2008AJ....135..850D}
{Devor}, J., {Charbonneau}, D., {O'Donovan}, F.~T., {Mandushev}, G., \& {Torres}, G. 2008, \aj, 135, 850

\bibitem[{{Donati} {et~al.}(2015){Donati}, {H{\'e}brard}, {Hussain}, {Moutou}, {Malo}, {Grankin}, {Vidotto}, {Alencar}, {Gregory}, {Jardine}, {Herczeg}, {Morin}, {Fares}, {M{\'e}nard}, {Bouvier}, {Delfosse}, {Doyon}, {Takami}, {Figueira}, {Petit}, {Boisse}, \& {MaTYSSE Collaboration}}]{2015MNRAS.453.3706D}
{Donati}, J.~F., {H{\'e}brard}, E., {Hussain}, G.~A.~J., {et~al.} 2015, \mnras, 453, 3706

\bibitem[{{Drake} {et~al.}(1987){Drake}, {Abbott}, {Bastian}, {Bieging}, {Churchwell}, {Dulk}, \& {Linsky}}]{1987ApJ...322..902D}
{Drake}, S.~A., {Abbott}, D.~C., {Bastian}, T.~S., {et~al.} 1987, \apj, 322, 902

\bibitem[{{Driessen} {et~al.}(2024){Driessen}, {Pritchard}, {Murphy}, {Heald}, {Robrade}, {Das}, {Duchesne}, {Kaplan}, {Lenc}, {Lynch}, {Mitchell-Bolton}, {Pope}, {Rose}, {Stelzer}, {Wang}, \& {Zic}}]{2024PASA...41...84D}
{Driessen}, L.~N., {Pritchard}, J., {Murphy}, T., {et~al.} 2024, \pasa, 41, e084

\bibitem[{{Dulk}(1985)}]{1985ARA&A..23..169D}
{Dulk}, G.~A. 1985, \araa, 23, 169

\bibitem[{{Dworetsky} {et~al.}(1982){Dworetsky}, {Whitelock}, \& {Carnochan}}]{1982MNRAS.201..901D}
{Dworetsky}, M.~M., {Whitelock}, P.~A., \& {Carnochan}, D.~J. 1982, \mnras, 201, 901

\bibitem[{{Fern{\'a}ndez} {et~al.}(2004){Fern{\'a}ndez}, {Stelzer}, {Henden}, {Grankin}, {Gameiro}, {Costa}, {Guenther}, {Amado}, \& {Rodriguez}}]{2004A&A...427..263F}
{Fern{\'a}ndez}, M., {Stelzer}, B., {Henden}, A., {et~al.} 2004, \aap, 427, 263

\bibitem[{{Fichtinger} {et~al.}(2017){Fichtinger}, {G{\"u}del}, {Mutel}, {Hallinan}, {Gaidos}, {Skinner}, {Lynch}, \& {Gayley}}]{2017A&A...599A.127F}
{Fichtinger}, B., {G{\"u}del}, M., {Mutel}, R.~L., {et~al.} 2017, \aap, 599, A127

\bibitem[{{Flower}(1996)}]{1996ApJ...469..355F}
{Flower}, P.~J. 1996, \apj, 469, 355

\bibitem[{{Frail} {et~al.}(2024){Frail}, {Polisensky}, {Hyman}, {Cotton}, {Kassim}, {Silverstein}, {Sengar}, {Kaplan}, {Calore}, {Berteaud}, {Clavel}, {Geyer}, {Legodi}, {Krishnan}, {Buchner}, \& {Camilo}}]{2024ApJ...975...34F}
{Frail}, D.~A., {Polisensky}, E., {Hyman}, S.~D., {et~al.} 2024, \apj, 975, 34

\bibitem[{{Frasca} {et~al.}(2018){Frasca}, {Guillout}, {Klutsch}, {Ferrero}, {Marilli}, {Biazzo}, {Gandolfi}, \& {Montes}}]{2018A&A...612A..96F}
{Frasca}, A., {Guillout}, P., {Klutsch}, A., {et~al.} 2018, \aap, 612, A96

\bibitem[{{Gaia Collaboration}(2022)}]{2022yCat.1358....0G}
{Gaia Collaboration}. 2022, {VizieR Online Data Catalog}, I/358

\bibitem[{{Gaia Collaboration} {et~al.}(2018){Gaia Collaboration}, {Brown}, {Vallenari}, {Prusti}, {de Bruijne}, {Babusiaux}, {Bailer-Jones}, {Biermann}, {Evans}, {Eyer}, {Jansen}, {Jordi}, {Klioner}, {Lammers}, {Lindegren}, {Luri}, {Mignard}, {Panem}, {Pourbaix}, {Randich}, {Sartoretti}, {Siddiqui}, {Soubiran}, {van Leeuwen}, {Walton}, {Arenou}, {Bastian}, {Cropper}, {Drimmel}, {Katz}, {Lattanzi}, {Bakker}, {Cacciari}, {Casta{\~n}eda}, {Chaoul}, {Cheek}, {De Angeli}, {Fabricius}, {Guerra}, {Holl}, {Masana}, {Messineo}, {Mowlavi}, {Nienartowicz}, {Panuzzo}, {Portell}, {Riello}, {Seabroke}, {Tanga}, {Th{\'e}venin}, {Gracia-Abril}, {Comoretto}, {Garcia-Reinaldos}, {Teyssier}, {Altmann}, {Andrae}, {Audard}, {Bellas-Velidis}, {Benson}, {Berthier}, {Blomme}, {Burgess}, {Busso}, {Carry}, {Cellino}, {Clementini}, {Clotet}, {Creevey}, {Davidson}, {De Ridder}, {Delchambre}, {Dell'Oro}, {Ducourant}, {Fern{\'a}ndez-Hern{\'a}ndez}, {Fouesneau}, {Fr{\'e}mat}, {Galluccio}, {Garc{\'\i}a-Torres},
  {Gonz{\'a}lez-N{\'u}{\~n}ez}, {Gonz{\'a}lez-Vidal}, {Gosset}, {Guy}, {Halbwachs}, {Hambly}, {Harrison}, {Hern{\'a}ndez}, {Hestroffer}, {Hodgkin}, {Hutton}, {Jasniewicz}, {Jean-Antoine-Piccolo}, {Jordan}, {Korn}, {Krone-Martins}, {Lanzafame}, {Lebzelter}, {L{\"o}ffler}, {Manteiga}, {Marrese}, {Mart{\'\i}n-Fleitas}, {Moitinho}, {Mora}, {Muinonen}, {Osinde}, {Pancino}, {Pauwels}, {Petit}, {Recio-Blanco}, {Richards}, {Rimoldini}, {Robin}, {Sarro}, {Siopis}, {Smith}, {Sozzetti}, {S{\"u}veges}, {Torra}, {van Reeven}, {Abbas}, {Abreu Aramburu}, {Accart}, {Aerts}, {Altavilla}, {{\'A}lvarez}, {Alvarez}, {Alves}, {Anderson}, {Andrei}, {Anglada Varela}, {Antiche}, {Antoja}, {Arcay}, {Astraatmadja}, {Bach}, {Baker}, {Balaguer-N{\'u}{\~n}ez}, {Balm}, {Barache}, {Barata}, {Barbato}, {Barblan}, {Barklem}, {Barrado}, {Barros}, {Barstow}, {Bartholom{\'e} Mu{\~n}oz}, {Bassilana}, {Becciani}, {Bellazzini}, {Berihuete}, {Bertone}, {Bianchi}, {Bienaym{\'e}}, {Blanco-Cuaresma}, {Boch}, {Boeche}, {Bombrun}, {Borrachero},
  {Bossini}, {Bouquillon}, {Bourda}, {Bragaglia}, {Bramante}, {Breddels}, {Bressan}, {Brouillet}, {Br{\"u}semeister}, {Brugaletta}, {Bucciarelli}, {Burlacu}, {Busonero}, {Butkevich}, {Buzzi}, {Caffau}, {Cancelliere}, {Cannizzaro}, {Cantat-Gaudin}, {Carballo}, {Carlucci}, {Carrasco}, {Casamiquela}, {Castellani}, {Castro-Ginard}, {Charlot}, {Chemin}, {Chiavassa}, {Cocozza}, {Costigan}, {Cowell}, {Crifo}, {Crosta}, {Crowley}, {Cuypers}, {Dafonte}, {Damerdji}, {Dapergolas}, {David}, {David}, {de Laverny}, \& {De Luise}}]{2018A&A...616A...1G}
{Gaia Collaboration}, {Brown}, A.~G.~A., {Vallenari}, A., {et~al.} 2018, \aap, 616, A1

\bibitem[{{Gaia Collaboration} {et~al.}(2016){Gaia Collaboration}, {Prusti}, {de Bruijne}, {Brown}, {Vallenari}, {Babusiaux}, {Bailer-Jones}, {Bastian}, {Biermann}, {Evans}, {Eyer}, {Jansen}, {Jordi}, {Klioner}, {Lammers}, {Lindegren}, {Luri}, {Mignard}, {Milligan}, {Panem}, {Poinsignon}, {Pourbaix}, {Randich}, {Sarri}, {Sartoretti}, {Siddiqui}, {Soubiran}, {Valette}, {van Leeuwen}, {Walton}, {Aerts}, {Arenou}, {Cropper}, {Drimmel}, {H{\o}g}, {Katz}, {Lattanzi}, {O'Mullane}, {Grebel}, {Holland}, {Huc}, {Passot}, {Bramante}, {Cacciari}, {Casta{\~n}eda}, {Chaoul}, {Cheek}, {De Angeli}, {Fabricius}, {Guerra}, {Hern{\'a}ndez}, {Jean-Antoine-Piccolo}, {Masana}, {Messineo}, {Mowlavi}, {Nienartowicz}, {Ord{\'o}{\~n}ez-Blanco}, {Panuzzo}, {Portell}, {Richards}, {Riello}, {Seabroke}, {Tanga}, {Th{\'e}venin}, {Torra}, {Els}, {Gracia-Abril}, {Comoretto}, {Garcia-Reinaldos}, {Lock}, {Mercier}, {Altmann}, {Andrae}, {Astraatmadja}, {Bellas-Velidis}, {Benson}, {Berthier}, {Blomme}, {Busso}, {Carry}, {Cellino}, {Clementini},
  {Cowell}, {Creevey}, {Cuypers}, {Davidson}, {De Ridder}, {de Torres}, {Delchambre}, {Dell'Oro}, {Ducourant}, {Fr{\'e}mat}, {Garc{\'\i}a-Torres}, {Gosset}, {Halbwachs}, {Hambly}, {Harrison}, {Hauser}, {Hestroffer}, {Hodgkin}, {Huckle}, {Hutton}, {Jasniewicz}, {Jordan}, {Kontizas}, {Korn}, {Lanzafame}, {Manteiga}, {Moitinho}, {Muinonen}, {Osinde}, {Pancino}, {Pauwels}, {Petit}, {Recio-Blanco}, {Robin}, {Sarro}, {Siopis}, {Smith}, {Smith}, {Sozzetti}, {Thuillot}, {van Reeven}, {Viala}, {Abbas}, {Abreu Aramburu}, {Accart}, {Aguado}, {Allan}, {Allasia}, {Altavilla}, {{\'A}lvarez}, {Alves}, {Anderson}, {Andrei}, {Anglada Varela}, {Antiche}, {Antoja}, {Ant{\'o}n}, {Arcay}, {Atzei}, {Ayache}, {Bach}, {Baker}, {Balaguer-N{\'u}{\~n}ez}, {Barache}, {Barata}, {Barbier}, {Barblan}, {Baroni}, {Barrado y Navascu{\'e}s}, {Barros}, {Barstow}, {Becciani}, {Bellazzini}, {Bellei}, {Bello Garc{\'\i}a}, {Belokurov}, {Bendjoya}, {Berihuete}, {Bianchi}, {Bienaym{\'e}}, {Billebaud}, {Blagorodnova}, {Blanco-Cuaresma}, {Boch},
  {Bombrun}, {Borrachero}, {Bouquillon}, {Bourda}, {Bouy}, {Bragaglia}, {Breddels}, {Brouillet}, {Br{\"u}semeister}, {Bucciarelli}, {Budnik}, {Burgess}, {Burgon}, {Burlacu}, {Busonero}, {Buzzi}, {Caffau}, {Cambras}, {Campbell}, {Cancelliere}, {Cantat-Gaudin}, {Carlucci}, {Carrasco}, {Castellani}, {Charlot}, {Charnas}, {Charvet}, {Chassat}, {Chiavassa}, {Clotet}, {Cocozza}, {Collins}, {Collins}, \& {Costigan}}]{2016A&A...595A...1G}
{Gaia Collaboration}, {Prusti}, T., {de Bruijne}, J.~H.~J., {et~al.} 2016, \aap, 595, A1

\bibitem[{{Gaia Collaboration} {et~al.}(2023){Gaia Collaboration}, {Vallenari}, {Brown}, {Prusti}, {de Bruijne}, {Arenou}, {Babusiaux}, {Biermann}, {Creevey}, {Ducourant}, {Evans}, {Eyer}, {Guerra}, {Hutton}, {Jordi}, {Klioner}, {Lammers}, {Lindegren}, {Luri}, {Mignard}, {Panem}, {Pourbaix}, {Randich}, {Sartoretti}, {Soubiran}, {Tanga}, {Walton}, {Bailer-Jones}, {Bastian}, {Drimmel}, {Jansen}, {Katz}, {Lattanzi}, {van Leeuwen}, {Bakker}, {Cacciari}, {Casta{\~n}eda}, {De Angeli}, {Fabricius}, {Fouesneau}, {Fr{\'e}mat}, {Galluccio}, {Guerrier}, {Heiter}, {Masana}, {Messineo}, {Mowlavi}, {Nicolas}, {Nienartowicz}, {Pailler}, {Panuzzo}, {Riclet}, {Roux}, {Seabroke}, {Sordo}, {Th{\'e}venin}, {Gracia-Abril}, {Portell}, {Teyssier}, {Altmann}, {Andrae}, {Audard}, {Bellas-Velidis}, {Benson}, {Berthier}, {Blomme}, {Burgess}, {Busonero}, {Busso}, {C{\'a}novas}, {Carry}, {Cellino}, {Cheek}, {Clementini}, {Damerdji}, {Davidson}, {de Teodoro}, {Nu{\~n}ez Campos}, {Delchambre}, {Dell'Oro}, {Esquej},
  {Fern{\'a}ndez-Hern{\'a}ndez}, {Fraile}, {Garabato}, {Garc{\'\i}a-Lario}, {Gosset}, {Haigron}, {Halbwachs}, {Hambly}, {Harrison}, {Hern{\'a}ndez}, {Hestroffer}, {Hodgkin}, {Holl}, {Jan{\ss}en}, {Jevardat de Fombelle}, {Jordan}, {Krone-Martins}, {Lanzafame}, {L{\"o}ffler}, {Marchal}, {Marrese}, {Moitinho}, {Muinonen}, {Osborne}, {Pancino}, {Pauwels}, {Recio-Blanco}, {Reyl{\'e}}, {Riello}, {Rimoldini}, {Roegiers}, {Rybizki}, {Sarro}, {Siopis}, {Smith}, {Sozzetti}, {Utrilla}, {van Leeuwen}, {Abbas}, {{\'A}brah{\'a}m}, {Abreu Aramburu}, {Aerts}, {Aguado}, {Ajaj}, {Aldea-Montero}, {Altavilla}, {{\'A}lvarez}, {Alves}, {Anders}, {Anderson}, {Anglada Varela}, {Antoja}, {Baines}, {Baker}, {Balaguer-N{\'u}{\~n}ez}, {Balbinot}, {Balog}, {Barache}, {Barbato}, {Barros}, {Barstow}, {Bartolom{\'e}}, {Bassilana}, {Bauchet}, {Becciani}, {Bellazzini}, {Berihuete}, {Bernet}, {Bertone}, {Bianchi}, {Binnenfeld}, {Blanco-Cuaresma}, {Blazere}, {Boch}, {Bombrun}, {Bossini}, {Bouquillon}, {Bragaglia}, {Bramante}, {Breedt},
  {Bressan}, {Brouillet}, {Brugaletta}, {Bucciarelli}, {Burlacu}, {Butkevich}, {Buzzi}, {Caffau}, {Cancelliere}, {Cantat-Gaudin}, {Carballo}, {Carlucci}, {Carnerero}, {Carrasco}, {Casamiquela}, {Castellani}, {Castro-Ginard}, {Chaoul}, {Charlot}, {Chemin}, {Chiaramida}, {Chiavassa}, {Chornay}, {Comoretto}, {Contursi}, {Cooper}, {Cornez}, {Cowell}, {Crifo}, {Cropper}, {Crosta}, {Crowley}, {Dafonte}, {Dapergolas}, {David}, {David}, {de Laverny}, {De Luise}, \& {De March}}]{2023A&A...674A...1G}
{Gaia Collaboration}, {Vallenari}, A., {Brown}, A.~G.~A., {et~al.} 2023, \aap, 674, A1

\bibitem[{{Galli} {et~al.}(2018){Galli}, {Loinard}, {Ortiz-L{\'e}on}, {Kounkel}, {Dzib}, {Mioduszewski}, {Rodr{\'\i}guez}, {Hartmann}, {Teixeira}, {Torres}, {Rivera}, {Boden}, {Evans}, {Brice{\~n}o}, {Tobin}, \& {Heyer}}]{2018ApJ...859...33G}
{Galli}, P. A.~B., {Loinard}, L., {Ortiz-L{\'e}on}, G.~N., {et~al.} 2018, \apj, 859, 33

\bibitem[{{Garc{\'\i}a} {et~al.}(2014){Garc{\'\i}a}, {Ceillier}, {Salabert}, {Mathur}, {van Saders}, {Pinsonneault}, {Ballot}, {Beck}, {Bloemen}, {Campante}, {Davies}, {do Nascimento}, {Mathis}, {Metcalfe}, {Nielsen}, {Su{\'a}rez}, {Chaplin}, {Jim{\'e}nez}, \& {Karoff}}]{2014A&A...572A..34G}
{Garc{\'\i}a}, R.~A., {Ceillier}, T., {Salabert}, D., {et~al.} 2014, \aap, 572, A34

\bibitem[{{Garc{\'\i}a-S{\'a}nchez} {et~al.}(2003){Garc{\'\i}a-S{\'a}nchez}, {Paredes}, \& {Rib{\'o}}}]{2003A&A...403..613G}
{Garc{\'\i}a-S{\'a}nchez}, J., {Paredes}, J.~M., \& {Rib{\'o}}, M. 2003, \aap, 403, 613

\bibitem[{{Ghez} {et~al.}(1993){Ghez}, {Neugebauer}, \& {Matthews}}]{1993AJ....106.2005G}
{Ghez}, A.~M., {Neugebauer}, G., \& {Matthews}, K. 1993, \aj, 106, 2005

\bibitem[{{Gordon} {et~al.}(2020){Gordon}, {Boyce}, {O'Dea}, {Rudnick}, {Andernach}, {Vantyghem}, {Baum}, {Bui}, \& {Dionyssiou}}]{2020RNAAS...4..175G}
{Gordon}, Y.~A., {Boyce}, M.~M., {O'Dea}, C.~P., {et~al.} 2020, Research Notes of the American Astronomical Society, 4, 175

\bibitem[{{Gordon} {et~al.}(2021){Gordon}, {Boyce}, {O'Dea}, {Rudnick}, {Andernach}, {Vantyghem}, {Baum}, {Bui}, {Dionyssiou}, {Safi-Harb}, \& {Sander}}]{2021ApJS..255...30G}
{Gordon}, Y.~A., {Boyce}, M.~M., {O'Dea}, C.~P., {et~al.} 2021, \apjs, 255, 30

\bibitem[{{Greisen}(2003)}]{2003ASSL..285..109G}
{Greisen}, E.~W. 2003, in Astrophysics and Space Science Library, Vol. 285, Information Handling in Astronomy - Historical Vistas, ed. A.~{Heck}, 109

\bibitem[{{Grie{\ss}meier} {et~al.}(2007){Grie{\ss}meier}, {Zarka}, \& {Spreeuw}}]{2007A&A...475..359G}
{Grie{\ss}meier}, J.~M., {Zarka}, P., \& {Spreeuw}, H. 2007, \aap, 475, 359

\bibitem[{{G{\"u}del}(2002)}]{2002ARA&A..40..217G}
{G{\"u}del}, M. 2002, \araa, 40, 217

\bibitem[{{G{\"u}del} \& {Benz}(1993)}]{1993ApJ...405L..63G}
{G{\"u}del}, M. \& {Benz}, A.~O. 1993, \apjl, 405, L63

\bibitem[{{Hallinan} {et~al.}(2015){Hallinan}, {Littlefair}, {Cotter}, {Bourke}, {Harding}, {Pineda}, {Butler}, {Golden}, {Basri}, {Doyle}, {Kao}, {Berdyugina}, {Kuznetsov}, {Rupen}, \& {Antonova}}]{2015Natur.523..568H}
{Hallinan}, G., {Littlefair}, S.~P., {Cotter}, G., {et~al.} 2015, \nat, 523, 568

\bibitem[{{Harmanec} {et~al.}(2015){Harmanec}, {Koubsk{\'y}}, {Nemravov{\'a}}, {Royer}, {Briot}, {North}, {Lampens}, {Fr{\'e}mat}, {Yang}, {Bo{\v{z}}i{\'c}}, {Kotkov{\'a}}, {{\v{S}}koda}, {{\v{S}}lechta}, {Kor{\v{c}}{\'a}kov{\'a}}, {Wolf}, \& {Zasche}}]{2015A&A...573A.107H}
{Harmanec}, P., {Koubsk{\'y}}, P., {Nemravov{\'a}}, J.~A., {et~al.} 2015, \aap, 573, A107

\bibitem[{{Harris} {et~al.}(2012){Harris}, {Andrews}, {Wilner}, \& {Kraus}}]{2012ApJ...751..115H}
{Harris}, R.~J., {Andrews}, S.~M., {Wilner}, D.~J., \& {Kraus}, A.~L. 2012, \apj, 751, 115

\bibitem[{H{\'{e}}der {et~al.}(2022)H{\'{e}}der, Rig{\'{o}}, Medgyesi, Lovas, Tenczer, Török, Farkas, Em{\H{o}}di, Kadlecsik, Mez{\H{o}}, Pint{\'{e}}r, \& Kacsuk}]{MTACloud}
H{\'{e}}der, M., Rig{\'{o}}, E., Medgyesi, D., {et~al.} 2022, Inform{\'{a}}ci{\'{o}}s T{\'{a}}rsadalom, 22, 128

\bibitem[{{Helfand} {et~al.}(2015){Helfand}, {White}, \& {Becker}}]{2015ApJ...801...26H}
{Helfand}, D.~J., {White}, R.~L., \& {Becker}, R.~H. 2015, \apj, 801, 26

\bibitem[{{Herbig}(1977)}]{1977ApJ...214..747H}
{Herbig}, G.~H. 1977, \apj, 214, 747

\bibitem[{{Herczeg} \& {Hillenbrand}(2014)}]{2014ApJ...786...97H}
{Herczeg}, G.~J. \& {Hillenbrand}, L.~A. 2014, \apj, 786, 97

\bibitem[{{Hotan} {et~al.}(2021){Hotan}, {Bunton}, {Chippendale}, {Whiting}, {Tuthill}, {Moss}, {McConnell}, {Amy}, {Huynh}, {Allison}, {Anderson}, {Bannister}, {Bastholm}, {Beresford}, {Bock}, {Bolton}, {Chapman}, {Chow}, {Collier}, {Cooray}, {Cornwell}, {Diamond}, {Edwards}, {Feain}, {Franzen}, {George}, {Gupta}, {Hampson}, {Harvey-Smith}, {Hayman}, {Heywood}, {Jacka}, {Jackson}, {Jackson}, {Jeganathan}, {Johnston}, {Kesteven}, {Kleiner}, {Koribalski}, {Lee-Waddell}, {Lenc}, {Lensson}, {Mackay}, {Mahony}, {McClure-Griffiths}, {McConigley}, {Mirtschin}, {Ng}, {Norris}, {Pearce}, {Phillips}, {Pilawa}, {Raja}, {Reynolds}, {Roberts}, {Roxby}, {Sadler}, {Shields}, {Schinckel}, {Serra}, {Shaw}, {Sweetnam}, {Troup}, {Tzioumis}, {Voronkov}, \& {Westmeier}}]{2021PASA...38....9H}
{Hotan}, A.~W., {Bunton}, J.~D., {Chippendale}, A.~P., {et~al.} 2021, \pasa, 38, e009

\bibitem[{Houk(1982)}]{1982MSS...C03....0H}
Houk, N. 1982, Michigan Catalogue of Two-dimensional Spectral Types for HD Stars: Declinations $-40\fdg$0 to $-26\fdg0$, Vol.~3 (University of Michigan)

\bibitem[{{Ivezi{\'c}} {et~al.}(2002){Ivezi{\'c}}, {Menou}, {Knapp}, {Strauss}, {Lupton}, {Vanden Berk}, {Richards}, {Tremonti}, {Weinstein}, {Anderson}, {Bahcall}, {Becker}, {Bernardi}, {Blanton}, {Eisenstein}, {Fan}, {Finkbeiner}, {Finlator}, {Frieman}, {Gunn}, {Hall}, {Kim}, {Kinkhabwala}, {Narayanan}, {Rockosi}, {Schlegel}, {Schneider}, {Strateva}, {SubbaRao}, {Thakar}, {Voges}, {White}, {Yanny}, {Brinkmann}, {Doi}, {Fukugita}, {Hennessy}, {Munn}, {Nichol}, \& {York}}]{2002AJ....124.2364I}
{Ivezi{\'c}}, {\v{Z}}., {Menou}, K., {Knapp}, G.~R., {et~al.} 2002, \aj, 124, 2364

\bibitem[{{Johnstone} {et~al.}(2015){Johnstone}, {G{\"u}del}, {Brott}, \& {L{\"u}ftinger}}]{2015A&A...577A..28J}
{Johnstone}, C.~P., {G{\"u}del}, M., {Brott}, I., \& {L{\"u}ftinger}, T. 2015, \aap, 577, A28

\bibitem[{{Joy} \& {Wilson}(1949)}]{1949ApJ...109..231J}
{Joy}, A.~H. \& {Wilson}, R.~E. 1949, \apj, 109, 231

\bibitem[{{Kao} {et~al.}(2023){Kao}, {Mioduszewski}, {Villadsen}, \& {Shkolnik}}]{2023Natur.619..272K}
{Kao}, M.~M., {Mioduszewski}, A.~J., {Villadsen}, J., \& {Shkolnik}, E.~L. 2023, \nat, 619, 272

\bibitem[{{Kavanagh} {et~al.}(2021){Kavanagh}, {Vidotto}, {Klein}, {Jardine}, {Donati}, \& {{\'O} Fionnag{\'a}in}}]{2021MNRAS.504.1511K}
{Kavanagh}, R.~D., {Vidotto}, A.~A., {Klein}, B., {et~al.} 2021, \mnras, 504, 1511

\bibitem[{{Keenan} \& {McNeil}(1989)}]{1989ApJS...71..245K}
{Keenan}, P.~C. \& {McNeil}, R.~C. 1989, \apjs, 71, 245

\bibitem[{{Kiraga}(2012)}]{2012AcA....62...67K}
{Kiraga}, M. 2012, \actaa, 62, 67

\bibitem[{{Kraus} {et~al.}(2017){Kraus}, {Herczeg}, {Rizzuto}, {Mann}, {Slesnick}, {Carpenter}, {Hillenbrand}, \& {Mamajek}}]{2017ApJ...838..150K}
{Kraus}, A.~L., {Herczeg}, G.~J., {Rizzuto}, A.~C., {et~al.} 2017, \apj, 838, 150

\bibitem[{{Kuijpers} \& {van der Hulst}(1985)}]{1985A&A...149..343K}
{Kuijpers}, J. \& {van der Hulst}, J.~M. 1985, \aap, 149, 343

\bibitem[{{Lef{\`e}vre} {et~al.}(1994){Lef{\`e}vre}, {Klein}, \& {Lestrade}}]{1994SSRv...68..293L}
{Lef{\`e}vre}, E., {Klein}, K.~L., \& {Lestrade}, J.~F. 1994, \ssr, 68, 293

\bibitem[{{Leinert} {et~al.}(1993){Leinert}, {Zinnecker}, {Weitzel}, {Christou}, {Ridgway}, {Jameson}, {Haas}, \& {Lenzen}}]{1993A&A...278..129L}
{Leinert}, C., {Zinnecker}, H., {Weitzel}, N., {et~al.} 1993, \aap, 278, 129

\bibitem[{{Lin} {et~al.}(2012){Lin}, {Webb}, \& {Barret}}]{2012ApJ...756...27L}
{Lin}, D., {Webb}, N.~A., \& {Barret}, D. 2012, \apj, 756, 27

\bibitem[{{Linsky} {et~al.}(1992){Linsky}, {Drake}, \& {Bastian}}]{1992ApJ...393..341L}
{Linsky}, J.~L., {Drake}, S.~A., \& {Bastian}, T.~S. 1992, \apj, 393, 341

\bibitem[{{Liu} {et~al.}(2021){Liu}, {Fang}, {Tian}, {Liu}, {Yang}, \& {Xue}}]{2021ApJS..254...20L}
{Liu}, J., {Fang}, M., {Tian}, H., {et~al.} 2021, \apjs, 254, 20

\bibitem[{{Lunz} {et~al.}(2023){Lunz}, {Anderson}, {Xu}, {Titov}, {Heinkelmann}, {Johnson}, \& {Schuh}}]{2023A&A...676A..11L}
{Lunz}, S., {Anderson}, J.~M., {Xu}, M.~H., {et~al.} 2023, \aap, 676, A11

\bibitem[{{Massi} {et~al.}(2008){Massi}, {Ros}, {Menten}, {Kaufman Bernad{\'o}}, {Torricelli-Ciamponi}, {Neidh{\"o}fer}, {Boden}, {Boboltz}, {Sargent}, \& {Torres}}]{2008A&A...480..489M}
{Massi}, M., {Ros}, E., {Menten}, K.~M., {et~al.} 2008, \aap, 480, 489

\bibitem[{{Matthews}(2019)}]{2019PASP..131a6001M}
{Matthews}, L.~D. 2019, \pasp, 131, 016001

\bibitem[{{Merloni} {et~al.}(2024){Merloni}, {Lamer}, {Liu}, {Ramos-Ceja}, {Brunner}, {Bulbul}, {Dennerl}, {Doroshenko}, {Freyberg}, {Friedrich}, {Gatuzz}, {Georgakakis}, {Haberl}, {Igo}, {Kreykenbohm}, {Liu}, {Maitra}, {Malyali}, {Mayer}, {Nandra}, {Predehl}, {Robrade}, {Salvato}, {Sanders}, {Stewart}, {Tub{\'\i}n-Arenas}, {Weber}, {Wilms}, {Arcodia}, {Artis}, {Aschersleben}, {Avakyan}, {Aydar}, {Bahar}, {Balzer}, {Becker}, {Berger}, {Boller}, {Bornemann}, {Br{\"u}ggen}, {Brusa}, {Buchner}, {Burwitz}, {Camilloni}, {Clerc}, {Comparat}, {Coutinho}, {Czesla}, {Dannhauer}, {Dauner}, {Dauser}, {Dietl}, {Dolag}, {Dwelly}, {Egg}, {Ehl}, {Freund}, {Friedrich}, {Gaida}, {Garrel}, {Ghirardini}, {Gokus}, {Gr{\"u}nwald}, {Grandis}, {Grotova}, {Gruen}, {Gueguen}, {H{\"a}mmerich}, {Hamaus}, {Hasinger}, {Haubner}, {Homan}, {Ider Chitham}, {Joseph}, {Joyce}, {K{\"o}nig}, {Kaltenbrunner}, {Khokhriakova}, {Kink}, {Kirsch}, {Kluge}, {Knies}, {Krippendorf}, {Krumpe}, {Kurpas}, {Li}, {Liu}, {Locatelli}, {Lorenz}, {M{\"u}ller},
  {Magaudda}, {Mannes}, {McCall}, {Meidinger}, {Michailidis}, {Migkas}, {Mu{\~n}oz-Giraldo}, {Musiimenta}, {Nguyen-Dang}, {Ni}, {Olechowska}, {Ota}, {Pacaud}, {Pasini}, {Perinati}, {Pires}, {Pommranz}, {Ponti}, {Poppenhaeger}, {P{\"u}hlhofer}, {Rau}, {Reh}, {Reiprich}, {Roster}, {Saeedi}, {Santangelo}, {Sasaki}, {Schmitt}, {Schneider}, {Schrabback}, {Schuster}, {Schwope}, {Seppi}, {Serim}, {Shreeram}, {Sokolova-Lapa}, {Starck}, {Stelzer}, {Stierhof}, {Suleimanov}, {Tenzer}, {Traulsen}, {Tr{\"u}mper}, {Tsuge}, {Urrutia}, {Veronica}, {Waddell}, {Willer}, {Wolf}, {Yeung}, {Zainab}, {Zangrandi}, {Zhang}, {Zhang}, \& {Zheng}}]{2024A&A...682A..34M}
{Merloni}, A., {Lamer}, G., {Liu}, T., {et~al.} 2024, \aap, 682, A34

\bibitem[{{Myers} {et~al.}(2015){Myers}, {Sande}, {Miller}, {Warren}, \& {Tracewell}}]{2015yCat.5145....0M}
{Myers}, J.~R., {Sande}, C.~B., {Miller}, A.~C., {Warren}, Jr., W.~H., \& {Tracewell}, D.~A. 2015, {VizieR Online Data Catalog: SKY2000 Master Catalog, Version 5 (Myers+ 2006)}, VizieR On-line Data Catalog: V/145. Originally published in: Goddard Space Flight Center, Flight Dynamics Division (2006)

\bibitem[{{Nesterov} {et~al.}(1995){Nesterov}, {Kuzmin}, {Ashimbaeva}, {Volchkov}, {R{\"o}ser}, \& {Bastian}}]{1995A&AS..110..367N}
{Nesterov}, V.~V., {Kuzmin}, A.~V., {Ashimbaeva}, N.~T., {et~al.} 1995, \aaps, 110, 367

\bibitem[{{Nguyen} {et~al.}(2012){Nguyen}, {Brandeker}, {van Kerkwijk}, \& {Jayawardhana}}]{2012ApJ...745..119N}
{Nguyen}, D.~C., {Brandeker}, A., {van Kerkwijk}, M.~H., \& {Jayawardhana}, R. 2012, \apj, 745, 119

\bibitem[{{Obonyo} {et~al.}(2024){Obonyo}, {Hoare}, {Lumsden}, {Thompson}, {Chibueze}, {Cotton}, {Rigby}, {Leto}, {Trigilio}, \& {Williams}}]{2024MNRAS.533.3862O}
{Obonyo}, W.~O., {Hoare}, M.~G., {Lumsden}, S.~L., {et~al.} 2024, \mnras, 533, 3862

\bibitem[{{Obonyo} {et~al.}(2019){Obonyo}, {Lumsden}, {Hoare}, {Purser}, {Kurtz}, \& {Johnston}}]{2019MNRAS.486.3664O}
{Obonyo}, W.~O., {Lumsden}, S.~L., {Hoare}, M.~G., {et~al.} 2019, \mnras, 486, 3664

\bibitem[{Ochsenbein(1996)}]{10.26093/cds/vizier}
Ochsenbein, F. 1996, The VizieR database of astronomical catalogues

\bibitem[{{Ochsenbein} {et~al.}(2000){Ochsenbein}, {Bauer}, \& {Marcout}}]{vizier2000}
{Ochsenbein}, F., {Bauer}, P., \& {Marcout}, J. 2000, \aaps, 143, 23

\bibitem[{{Park} {et~al.}(2021){Park}, {Lee}, {Contreras Pe{\~n}a}, {Johnstone}, {Herczeg}, {Lee}, {Lee}, {Bhardwaj}, \& {Moriarty-Schieven}}]{2021ApJ...920..132P}
{Park}, W., {Lee}, J.-E., {Contreras Pe{\~n}a}, C., {et~al.} 2021, \apj, 920, 132

\bibitem[{{Patterer} {et~al.}(1993){Patterer}, {Ramsey}, {Welty}, \& {Huenemoerder}}]{1993AJ....105.1519P}
{Patterer}, R.~J., {Ramsey}, L., {Welty}, A.~D., \& {Huenemoerder}, D.~P. 1993, \aj, 105, 1519

\bibitem[{{Pecaut} \& {Mamajek}(2016)}]{2016MNRAS.461..794P}
{Pecaut}, M.~J. \& {Mamajek}, E.~E. 2016, \mnras, 461, 794

\bibitem[{{Pietka} {et~al.}(2015){Pietka}, {Fender}, \& {Keane}}]{2015MNRAS.446.3687P}
{Pietka}, M., {Fender}, R.~P., \& {Keane}, E.~F. 2015, \mnras, 446, 3687

\bibitem[{{Pizzocaro} {et~al.}(2019){Pizzocaro}, {Stelzer}, {Poretti}, {Raetz}, {Micela}, {Belfiore}, {Marelli}, {Salvetti}, \& {De Luca}}]{2019A&A...628A..41P}
{Pizzocaro}, D., {Stelzer}, B., {Poretti}, E., {et~al.} 2019, \aap, 628, A41

\bibitem[{{Pizzolato} {et~al.}(2003){Pizzolato}, {Maggio}, {Micela}, {Sciortino}, \& {Ventura}}]{2003A&A...397..147P}
{Pizzolato}, N., {Maggio}, A., {Micela}, G., {Sciortino}, S., \& {Ventura}, P. 2003, \aap, 397, 147

\bibitem[{{Pope} {et~al.}(2021){Pope}, {Callingham}, {Feinstein}, {G{\"u}nther}, {Vedantham}, {Ansdell}, \& {Shimwell}}]{2021ApJ...919L..10P}
{Pope}, B. J.~S., {Callingham}, J.~R., {Feinstein}, A.~D., {et~al.} 2021, \apjl, 919, L10

\bibitem[{{Rahman} {et~al.}(2025){Rahman}, {Lovell}, {Koch}, {Wilner}, {Andrews}, {Monsch}, \& {Ha}}]{2025ApJ...981...68R}
{Rahman}, R.~A., {Lovell}, J.~B., {Koch}, E.~W., {et~al.} 2025, \apj, 981, 68

\bibitem[{{Reyl{\'e}}(2018)}]{2018A&A...619L...8R}
{Reyl{\'e}}, C. 2018, \aap, 619, L8

\bibitem[{{Riaz} {et~al.}(2006){Riaz}, {Gizis}, \& {Harvin}}]{2006AJ....132..866R}
{Riaz}, B., {Gizis}, J.~E., \& {Harvin}, J. 2006, \aj, 132, 866

\bibitem[{{Riviere-Marichalar} {et~al.}(2012){Riviere-Marichalar}, {M{\'e}nard}, {Thi}, {Kamp}, {Montesinos}, {Meeus}, {Woitke}, {Howard}, {Sandell}, {Podio}, {Dent}, {Mendigut{\'\i}a}, {Pinte}, {White}, \& {Barrado}}]{2012A&A...538L...3R}
{Riviere-Marichalar}, P., {M{\'e}nard}, F., {Thi}, W.~F., {et~al.} 2012, \aap, 538, L3

\bibitem[{{Roeser} \& {Bastian}(1988)}]{1988A&AS...74..449R}
{Roeser}, S. \& {Bastian}, U. 1988, \aaps, 74, 449

\bibitem[{{Rose} {et~al.}(2023){Rose}, {Pritchard}, {Murphy}, {Caleb}, {Dobie}, {Driessen}, {Duchesne}, {Kaplan}, {Lenc}, \& {Wang}}]{2023ApJ...951L..43R}
{Rose}, K., {Pritchard}, J., {Murphy}, T., {et~al.} 2023, \apjl, 951, L43

\bibitem[{{Rosen} {et~al.}(2016){Rosen}, {Webb}, {Watson}, {Ballet}, {Barret}, {Braito}, {Carrera}, {Ceballos}, {Coriat}, {Della Ceca}, {Denkinson}, {Esquej}, {Farrell}, {Freyberg}, {Gris{\'e}}, {Guillout}, {Heil}, {Koliopanos}, {Law-Green}, {Lamer}, {Lin}, {Martino}, {Michel}, {Motch}, {Nebot Gomez-Moran}, {Page}, {Page}, {Page}, {Pakull}, {Pye}, {Read}, {Rodriguez}, {Sakano}, {Saxton}, {Schwope}, {Scott}, {Sturm}, {Traulsen}, {Yershov}, \& {Zolotukhin}}]{2016A&A...590A...1R}
{Rosen}, S.~R., {Webb}, N.~A., {Watson}, M.~G., {et~al.} 2016, \aap, 590, A1

\bibitem[{{Royer} {et~al.}(2007){Royer}, {Zorec}, \& {G{\'o}mez}}]{2007A&A...463..671R}
{Royer}, F., {Zorec}, J., \& {G{\'o}mez}, A.~E. 2007, \aap, 463, 671

\bibitem[{{Samus'} {et~al.}(2003){Samus'}, {Goranskii}, {Durlevich}, {Zharova}, {Kazarovets}, {Kireeva}, {Pastukhova}, {Williams}, \& {Hazen}}]{2003AstL...29..468S}
{Samus'}, N.~N., {Goranskii}, V.~P., {Durlevich}, O.~V., {et~al.} 2003, Astronomy Letters, 29, 468

\bibitem[{{Seli} {et~al.}(2025){Seli}, {Vida}, {Ol{\'a}h}, {G{\"o}rgei}, {So{\'o}s}, {P{\'a}l}, {Kriskovics}, \& {K{\H{o}}v{\'a}ri}}]{2025A&A...694A.161S}
{Seli}, B., {Vida}, K., {Ol{\'a}h}, K., {et~al.} 2025, \aap, 694, A161

\bibitem[{{Stassun} {et~al.}(2019){Stassun}, {Oelkers}, {Paegert}, {Torres}, {Pepper}, {De Lee}, {Collins}, {Latham}, {Muirhead}, {Chittidi}, {Rojas-Ayala}, {Fleming}, {Rose}, {Tenenbaum}, {Ting}, {Kane}, {Barclay}, {Bean}, {Brassuer}, {Charbonneau}, {Ge}, {Lissauer}, {Mann}, {McLean}, {Mullally}, {Narita}, {Plavchan}, {Ricker}, {Sasselov}, {Seager}, {Sharma}, {Shiao}, {Sozzetti}, {Stello}, {Vanderspek}, {Wallace}, \& {Winn}}]{2019AJ....158..138S}
{Stassun}, K.~G., {Oelkers}, R.~J., {Paegert}, M., {et~al.} 2019, \aj, 158, 138

\bibitem[{{Strassmeier}(2009)}]{2009A&ARv..17..251S}
{Strassmeier}, K.~G. 2009, \aapr, 17, 251

\bibitem[{{Strassmeier} \& {Fekel}(1990)}]{1990A&A...230..389S}
{Strassmeier}, K.~G. \& {Fekel}, F.~C. 1990, \aap, 230, 389

\bibitem[{{Toet} {et~al.}(2021){Toet}, {Vedantham}, {Callingham}, {Veken}, {Shimwell}, {Zarka}, {R{\"o}ttgering}, \& {Drabent}}]{2021A&A...654A..21T}
{Toet}, S.~E.~B., {Vedantham}, H.~K., {Callingham}, J.~R., {et~al.} 2021, \aap, 654, A21

\bibitem[{{Torres} {et~al.}(2006){Torres}, {Quast}, {da Silva}, {de La Reza}, {Melo}, \& {Sterzik}}]{2006A&A...460..695T}
{Torres}, C.~A.~O., {Quast}, G.~R., {da Silva}, L., {et~al.} 2006, \aap, 460, 695

\bibitem[{{Torres} {et~al.}(2012){Torres}, {Loinard}, {Mioduszewski}, {Boden}, {Franco-Hern{\'a}ndez}, {Vlemmings}, \& {Rodr{\'\i}guez}}]{2012ApJ...747...18T}
{Torres}, R.~M., {Loinard}, L., {Mioduszewski}, A.~J., {et~al.} 2012, \apj, 747, 18

\bibitem[{{Trigilio} {et~al.}(2000){Trigilio}, {Leto}, {Leone}, {Umana}, \& {Buemi}}]{2000A&A...362..281T}
{Trigilio}, C., {Leto}, P., {Leone}, F., {Umana}, G., \& {Buemi}, C. 2000, \aap, 362, 281

\bibitem[{{Trigilio} {et~al.}(2018){Trigilio}, {Umana}, {Cavallaro}, {Agliozzo}, {Leto}, {Buemi}, {Ingallinera}, {Bufano}, \& {Riggi}}]{2018MNRAS.481..217T}
{Trigilio}, C., {Umana}, G., {Cavallaro}, F., {et~al.} 2018, \mnras, 481, 217

\bibitem[{{Umana} {et~al.}(1998){Umana}, {Trigilio}, \& {Catalano}}]{1998A&A...329.1010U}
{Umana}, G., {Trigilio}, C., \& {Catalano}, S. 1998, \aap, 329, 1010

\bibitem[{{van den Oord} \& {Doyle}(1997)}]{1997A&A...319..578V}
{van den Oord}, G.~H.~J. \& {Doyle}, J.~G. 1997, \aap, 319, 578

\bibitem[{{van Haarlem} {et~al.}(2013){van Haarlem}, {Wise}, {Gunst}, {Heald}, {McKean}, {Hessels}, {de Bruyn}, {Nijboer}, {Swinbank}, {Fallows}, {Brentjens}, {Nelles}, {Beck}, {Falcke}, {Fender}, {H{\"o}randel}, {Koopmans}, {Mann}, {Miley}, {R{\"o}ttgering}, {Stappers}, {Wijers}, {Zaroubi}, {van den Akker}, {Alexov}, {Anderson}, {Anderson}, {van Ardenne}, {Arts}, {Asgekar}, {Avruch}, {Batejat}, {B{\"a}hren}, {Bell}, {Bell}, {van Bemmel}, {Bennema}, {Bentum}, {Bernardi}, {Best}, {B{\^\i}rzan}, {Bonafede}, {Boonstra}, {Braun}, {Bregman}, {Breitling}, {van de Brink}, {Broderick}, {Broekema}, {Brouw}, {Br{\"u}ggen}, {Butcher}, {van Cappellen}, {Ciardi}, {Coenen}, {Conway}, {Coolen}, {Corstanje}, {Damstra}, {Davies}, {Deller}, {Dettmar}, {van Diepen}, {Dijkstra}, {Donker}, {Doorduin}, {Dromer}, {Drost}, {van Duin}, {Eisl{\"o}ffel}, {van Enst}, {Ferrari}, {Frieswijk}, {Gankema}, {Garrett}, {de Gasperin}, {Gerbers}, {de Geus}, {Grie{\ss}meier}, {Grit}, {Gruppen}, {Hamaker}, {Hassall}, {Hoeft}, {Holties},
  {Horneffer}, {van der Horst}, {van Houwelingen}, {Huijgen}, {Iacobelli}, {Intema}, {Jackson}, {Jelic}, {de Jong}, {Juette}, {Kant}, {Karastergiou}, {Koers}, {Kollen}, {Kondratiev}, {Kooistra}, {Koopman}, {Koster}, {Kuniyoshi}, {Kramer}, {Kuper}, {Lambropoulos}, {Law}, {van Leeuwen}, {Lemaitre}, {Loose}, {Maat}, {Macario}, {Markoff}, {Masters}, {McFadden}, {McKay-Bukowski}, {Meijering}, {Meulman}, {Mevius}, {Middelberg}, {Millenaar}, {Miller-Jones}, {Mohan}, {Mol}, {Morawietz}, {Morganti}, {Mulcahy}, {Mulder}, {Munk}, {Nieuwenhuis}, {van Nieuwpoort}, {Noordam}, {Norden}, {Noutsos}, {Offringa}, {Olofsson}, {Omar}, {Orr{\'u}}, {Overeem}, {Paas}, {Pandey-Pommier}, {Pandey}, {Pizzo}, {Polatidis}, {Rafferty}, {Rawlings}, {Reich}, {de Reijer}, {Reitsma}, {Renting}, {Riemers}, {Rol}, {Romein}, {Roosjen}, {Ruiter}, {Scaife}, {van der Schaaf}, {Scheers}, {Schellart}, {Schoenmakers}, {Schoonderbeek}, {Serylak}, {Shulevski}, {Sluman}, {Smirnov}, {Sobey}, {Spreeuw}, {Steinmetz}, {Sterks}, {Stiepel}, {Stuurwold},
  {Tagger}, {Tang}, {Tasse}, {Thomas}, {Thoudam}, {Toribio}, {van der Tol}, {Usov}, {van Veelen}, {van der Veen}, {ter Veen}, {Verbiest}, {Vermeulen}, {Vermaas}, {Vocks}, {Vogt}, {de Vos}, {van der Wal}, {van Weeren}, {Weggemans}, {Weltevrede}, {White}, {Wijnholds}, {Wilhelmsson}, {Wucknitz}, {Yatawatta}, {Zarka}, \& {Zensus}}]{2013A&A...556A...2V}
{van Haarlem}, M.~P., {Wise}, M.~W., {Gunst}, A.~W., {et~al.} 2013, \aap, 556, A2

\bibitem[{{Vedantham} {et~al.}(2022){Vedantham}, {Callingham}, {Shimwell}, {Benz}, {Hajduk}, {Ray}, {Tasse}, \& {Drabent}}]{2022ApJ...926L..30V}
{Vedantham}, H.~K., {Callingham}, J.~R., {Shimwell}, T.~W., {et~al.} 2022, \apjl, 926, L30

\bibitem[{{Vedantham} {et~al.}(2020){Vedantham}, {Callingham}, {Shimwell}, {Tasse}, {Pope}, {Bedell}, {Snellen}, {Best}, {Hardcastle}, {Haverkorn}, {Mechev}, {O'Sullivan}, {R{\"o}ttgering}, \& {White}}]{2020NatAs...4..577V}
{Vedantham}, H.~K., {Callingham}, J.~R., {Shimwell}, T.~W., {et~al.} 2020, Nature Astronomy, 4, 577

\bibitem[{{Vedantham} {et~al.}(2023){Vedantham}, {Dupuy}, {Evans}, {Sanghi}, {Callingham}, {Shimwell}, {Best}, {Liu}, \& {Zarka}}]{2023A&A...675L...6V}
{Vedantham}, H.~K., {Dupuy}, T.~J., {Evans}, E.~L., {et~al.} 2023, \aap, 675, L6

\bibitem[{{Vida} {et~al.}(2021){Vida}, {B{\'o}di}, {Szklen{\'a}r}, \& {Seli}}]{flatwrm2}
{Vida}, K., {B{\'o}di}, A., {Szklen{\'a}r}, T., \& {Seli}, B. 2021, \aap, 652, A107

\bibitem[{{Vida} {et~al.}(2024){Vida}, {K{\H{o}}v{\'a}ri}, {Leitzinger}, {Odert}, {Ol{\'a}h}, {Seli}, {Kriskovics}, {Greimel}, \& {G{\"o}rgei}}]{2024Univ...10..313V}
{Vida}, K., {K{\H{o}}v{\'a}ri}, {\relax Zs}., {Leitzinger}, M., {et~al.} 2024, Universe, 10, 313

\bibitem[{{Vida} {et~al.}(2014){Vida}, {Ol{\'a}h}, \& {Szab{\'o}}}]{2014MNRAS.441.2744V}
{Vida}, K., {Ol{\'a}h}, K., \& {Szab{\'o}}, R. 2014, \mnras, 441, 2744

\bibitem[{{Villadsen}(2017)}]{2017PhDT.........8V}
{Villadsen}, J.~R. 2017, PhD thesis, California Institute of Technology

\bibitem[{{Voges} {et~al.}(1999){Voges}, {Aschenbach}, {Boller}, {Br{\"a}uninger}, {Briel}, {Burkert}, {Dennerl}, {Englhauser}, {Gruber}, {Haberl}, {Hartner}, {Hasinger}, {K{\"u}rster}, {Pfeffermann}, {Pietsch}, {Predehl}, {Rosso}, {Schmitt}, {Tr{\"u}mper}, \& {Zimmermann}}]{1999A&A...349..389V}
{Voges}, W., {Aschenbach}, B., {Boller}, T., {et~al.} 1999, \aap, 349, 389

\bibitem[{{Voges} {et~al.}(2000){Voges}, {Aschenbach}, {Boller}, {Brauninger}, {Briel}, {Burkert}, {Dennerl}, {Englhauser}, {Gruber}, {Haberl}, {Hartner}, {Hasinger}, {Pfeffermann}, {Pietsch}, {Predehl}, {Schmitt}, {Trumper}, \& {Zimmermann}}]{2000IAUC.7432....3V}
{Voges}, W., {Aschenbach}, B., {Boller}, T., {et~al.} 2000, \iaucirc, 7432, 3

\bibitem[{{Wheatley} {et~al.}(2017){Wheatley}, {Louden}, {Bourrier}, {Ehrenreich}, \& {Gillon}}]{2017MNRAS.465L..74W}
{Wheatley}, P.~J., {Louden}, T., {Bourrier}, V., {Ehrenreich}, D., \& {Gillon}, M. 2017, \mnras, 465, L74

\bibitem[{{White} {et~al.}(2020){White}, {K{\'o}sp{\'a}l}, {Hughes}, {{\'A}brah{\'a}m}, {Akimkin}, {Banzatti}, {Chen}, {Cruz-S{\'a}enz de Miera}, {Dutrey}, {Flock}, {Guilloteau}, {Hales}, {Henning}, {Kadam}, {Semenov}, {Sicilia-Aguilar}, {Teague}, \& {Vorobyov}}]{2020ApJ...904...37W}
{White}, J.~A., {K{\'o}sp{\'a}l}, {\'A}., {Hughes}, A.~G., {et~al.} 2020, \apj, 904, 37

\bibitem[{{White} {et~al.}(2007){White}, {Gabor}, \& {Hillenbrand}}]{2007AJ....133.2524W}
{White}, R.~J., {Gabor}, J.~M., \& {Hillenbrand}, L.~A. 2007, \aj, 133, 2524

\bibitem[{{White} {et~al.}(1997){White}, {Becker}, {Helfand}, \& {Gregg}}]{1997ApJ...475..479W}
{White}, R.~L., {Becker}, R.~H., {Helfand}, D.~J., \& {Gregg}, M.~D. 1997, \apj, 475, 479

\bibitem[{{Williams} {et~al.}(2014){Williams}, {Cook}, \& {Berger}}]{2014ApJ...785....9W}
{Williams}, P.~K.~G., {Cook}, B.~A., \& {Berger}, E. 2014, \apj, 785, 9

\bibitem[{{Wright} {et~al.}(2011){Wright}, {Drake}, {Mamajek}, \& {Henry}}]{2011ApJ...743...48W}
{Wright}, N.~J., {Drake}, J.~J., {Mamajek}, E.~E., \& {Henry}, G.~W. 2011, \apj, 743, 48

\bibitem[{{Xiao} {et~al.}(2012){Xiao}, {Covey}, {Rebull}, {Charbonneau}, {Mandushev}, {O'Donovan}, {Slesnick}, \& {Lloyd}}]{2012ApJS..202....7X}
{Xiao}, H.~Y., {Covey}, K.~R., {Rebull}, L., {et~al.} 2012, \apjs, 202, 7

\bibitem[{{Zic} {et~al.}(2020){Zic}, {Murphy}, {Lynch}, {Heald}, {Lenc}, {Kaplan}, {Cairns}, {Coward}, {Gendre}, {Johnston}, {MacGregor}, {Price}, \& {Wheatland}}]{2020ApJ...905...23Z}
{Zic}, A., {Murphy}, T., {Lynch}, C., {et~al.} 2020, \apj, 905, 23

\end{thebibliography}

\begin{appendix}
\onecolumn
\section{Radio counterparts of flaring stars}\label{sec:radio_counterpart_app}
The final list of flaring stars with associated radio sources are listed in Table~\ref{tab:radiox}. The associations were defined to have SNR $\le 6$ intensity peaks in the VLASS data within an angular separation of $\sqrt{2}\arcsec$, with respect to the proper-motion corrected optical positions at the time of the VLASS observations. For the first and second epoch VLASS observations, flux densities are obtained from the respective VLASS component catalogues, while values for the third epoch were acquired from fitting elliptical Gaussian model components to the fits images in \code{aips}. For a non-detection of a source at a certain epoch, an upper limit is given instead of the flux density value, defined as $5\sigma$ rms image noise level.

\begin{figure*}[hbp]
    \centering
    \includegraphics[width=0.99\linewidth]{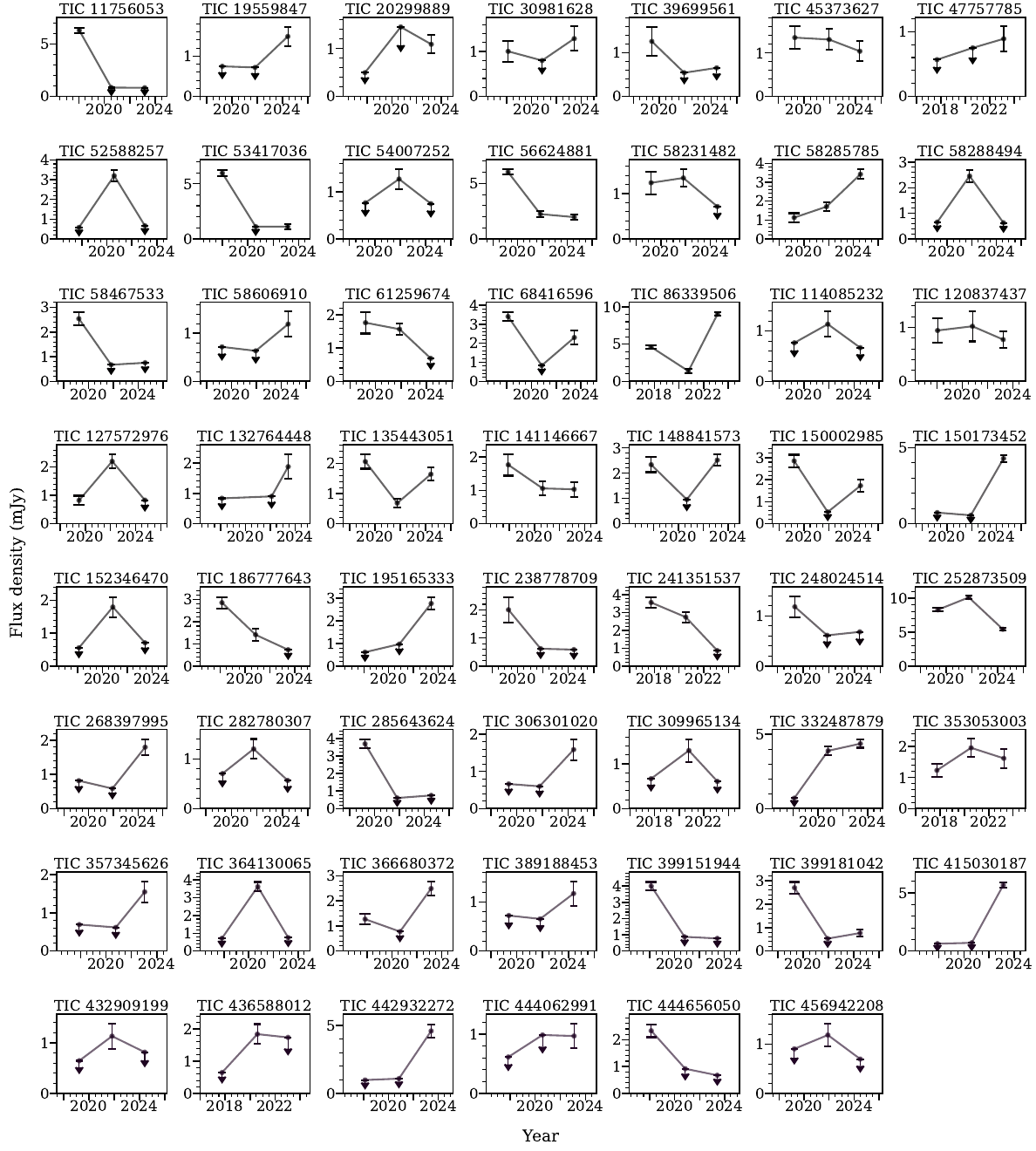}
    \caption{Radio light curves of flaring stars detected at 3~GHz in the VLASS. In the case of non-detection at a certain VLASS observing epoch, $5\sigma$ rms noise levels are given as upper limits. }
    \label{fig:radio-lc}
\end{figure*}

\begin{landscape}\footnotesize\setlength{\tabcolsep}{3.5pt}
\begin{longtable}{llccccrrrrrrccccc} 
\caption{List of the $55$ radio detected flaring stars. The 7 objects with coincident \textit{TESS} observations are marked with asterisks.}\label{tab:radiox}\\\hline\hline
\multirow{2}{*}{ID}	&	\multirow{2}{*}{Rate}	&	$T_\mathrm{eff}$			&	$P_\mathrm{rot}$			&	$T_\mathrm{gyro}$					&	\multicolumn{1}{c}{Spectral}	&	\multirow{2}{*}{$\mathrm{SNR}_{1}$}	&	\multicolumn{1}{c}{$S_{1}$}			&	\multirow{2}{*}{$\mathrm{SNR}_{2}$}	&	\multicolumn{1}{c}{$S_{2}$}			&	\multirow{2}{*}{$\mathrm{SNR}_{3}$}	&		\multicolumn{1}{c}{$S_{3}$}		&	$l_\mathrm{1RX}$	&	$l_\mathrm{2RX,PL}$	&	$l_\mathrm{eR,1}$	&	$l_\mathrm{eR,P3}$	&	$l_\mathrm{XMM}$	\\
	&		&	(K)			&	(day)			&	(Myr)					&	\multicolumn{1}{c}{type}	&		&	\multicolumn{1}{c}{(mJy)}			&		&	\multicolumn{1}{c}{(mJy)}			&		&		\multicolumn{1}{c}{(mJy)}		&	(erg~s$^{-1}$)	&	(erg~s$^{-1}$)	&	(erg~s$^{-1}$)	&	(erg~s$^{-1}$)	&	(erg~s$^{-1}$)	\\ \hline
\endfirsthead
\caption{continued.}\\
\hline\hline
\multirow{2}{*}{ID}	&	\multirow{2}{*}{Rate}	&	$T_\mathrm{eff}$			&	$P_\mathrm{rot}$			&	$T_\mathrm{gyro}$					&	\multicolumn{1}{c}{Spectral}	&	\multirow{2}{*}{$\mathrm{SNR}_{1}$}	&	\multicolumn{1}{c}{$S_{1}$}			&	\multirow{2}{*}{$\mathrm{SNR}_{2}$}	&	\multicolumn{1}{c}{$S_{2}$}			&	\multirow{2}{*}{$\mathrm{SNR}_{3}$}	&		\multicolumn{1}{c}{$S_{3}$}		&	$l_\mathrm{1RX}$	&	$l_\mathrm{2RX,PL}$	&	$l_\mathrm{eR,1}$	&	$l_\mathrm{eR,P3}$	&	$l_\mathrm{XMM}$	\\
	&		&	(K)			&	(day)			&	(Myr)					&	\multicolumn{1}{c}{type}	&		&	\multicolumn{1}{c}{(mJy)}			&		&	\multicolumn{1}{c}{(mJy)}			&		&		\multicolumn{1}{c}{(mJy)}		&	(erg~s$^{-1}$)	&	(erg~s$^{-1}$)	&	(erg~s$^{-1}$)	&	(erg~s$^{-1}$)	&	(erg~s$^{-1}$)	\\ \hline

\hline\endhead
\hline\endfoot
\hline\endlastfoot
TIC~11756053	&	0.538	&	4852	$\pm$	133	&	3.89	$\pm$	0.58	&	95	$^{+	200}	_{-	66}$	&	K0-2IV (15)	&	20.6	&	6.3	$\pm$	0.3	&	3.4	&	<0.9	  		&	3.3	&	<0.8	  		&	31.2	&	31.3	&	31.1	&	30.7	&		\\
TIC~19559847	&		&		  		&	1.64	$\pm$	0.25	&		  	 	  	 	&	K7 (26)	&	3.0	&	<0.7	  		&	3.5	&	<0.7	  		&	9.5	&	1.5	$\pm$	0.2	&	30.0	&	30.1	&	30.2	&	29.6	&		\\
TIC~20299889	&	0.090	&	4906	$\pm$	133	&		  		&		  	 	  	 	&	G8V (1)	&	3.9	&	<0.6	  		&	5.5	&	<0.7	  		&	9.8	&	1.1	$\pm$	0.2	&	30.7	&	31.3	&		&		&		\\
TIC~30981628	&	0.262	&	5038	$\pm$	124	&	3.41	$\pm$	0.51	&	90	$^{+	186}	_{-	62}$	&	G9III (11)	&	7.0	&	1.0	$\pm$	0.2	&	3.6	&	<0.7	  		&	7.5	&	1.3	$\pm$	0.3	&	30.9	&	31.6	&		&		&		\\
TIC~39699561	&	0.445	&	5651	$\pm$	125	&	5.69	$\pm$	0.85	&	321	$^{+	136}	_{-	149}$	&	G0 (9)	&	6.6	&	1.3	$\pm$	0.3	&	3.9	&	<0.6	  		&	3.4	&	<0.6	  		&	30.7	&	30.6	&		&		&		\\
TIC~45373627	&	0.136	&	5358	$\pm$	138	&	0.37	$\pm$	0.06	&	81	$^{+	159}	_{-	56}$	&	G5V (24)	&	7.7	&	1.4	$\pm$	0.3	&	8.4	&	1.3	$\pm$	0.2	&	7.2	&	1.0	$\pm$	0.2	&	30.6	&	31.2	&	30.7	&	30.1	&		\\
TIC~47757785	&	0.170	&	5581	$\pm$	124	&		  		&		  	 	  	 	&	G5 (9)	&	4.7	&	<0.6	  		&	3.7	&	<0.7	  		&	6.8	&	0.9	$\pm$	0.2	&	30.8	&	30.9	&		&		&		\\
TIC~52588257	&	0.139	&	5362	$\pm$	138	&	1.51	$\pm$	0.23	&	81	$^{+	159}	_{-	56}$	&	G0V (14)	&	3.5	&	<0.6	  		&	14.1	&	3.2	$\pm$	0.3	&	3.7	&	<0.7	  		&	30.7	&	30.8	&	30.6	&	30.1	&		\\
TIC~53417036	&	0.247	&	5555	$\pm$	109	&	1.53	$\pm$	0.23	&	77	$^{+	142}	_{-	53}$	&	G0 (10)	&	19.7	&	6.0	$\pm$	0.3	&	4.0	&	<1.0	  		&	6.5	&	1.1	$\pm$	0.2	&	30.3	&	30.4	&	30.4	&	29.9	&		\\
TIC~54007252$^*$	&	0.298	&	4489	$\pm$	124	&	2.48	$\pm$	0.37	&	96	$^{+	220}	_{-	67}$	&	G9 (17)	&	3.6	&	<0.7	  		&	9.6	&	1.3	$\pm$	0.2	&	3.5	&	<0.8	  		&	30.8	&		&	30.8	&	30.5	&		\\
TIC~56624881$^*$	&	0.317	&	4464	$\pm$	36	&	2.64	$\pm$	0.4	&	95	$^{+	221}	_{-	66}$	&	K3Ve (2)	&	20.8	&	6.0	$\pm$	0.2	&	12.3	&	2.2	$\pm$	0.3	&	12.9	&	2.0	$\pm$	0.2	&	30.6	&		&		&		&	30.3	\\
TIC~58231482$^*$	&	0.365	&	4329	$\pm$	131	&	1.87	$\pm$	0.28	&	100	$^{+	229}	_{-	70}$	&	K3Ve (2)	&	7.7	&	1.2	$\pm$	0.3	&	10.6	&	1.3	$\pm$	0.2	&	5.6	&	<0.7	  		&	30.6	&	30.5	&		&		&	30.2	\\
TIC~58285785$^*$	&	0.045	&	3637	$\pm$	129	&	1.56	$\pm$	0.23	&		  	 	  	 	&	K7 (19)	&	7.9	&	1.1	$\pm$	0.2	&	11.1	&	1.7	$\pm$	0.2	&	18.3	&	3.4	$\pm$	0.3	&	30.2	&		&		&		&	29.9	\\
TIC~58288494$^*$	&	0.118	&	3935	$\pm$	105	&	5.45	$\pm$	0.82	&	144	$^{+	245}	_{-	98}$	&	K7 (16)	&	3.5	&	<0.6	  		&	14.9	&	2.5	$\pm$	0.2	&	3.3	&	<0.6	  		&		&		&		&		&	29.6	\\
TIC~58467533	&	0.263	&	5857	$\pm$	113	&	1.53	$\pm$	0.23	&	66	$^{+	70}	_{-	45}$	&	G5IVe (8)	&	13.5	&	2.5	$\pm$	0.3	&	4.9	&	<0.6	  		&	4.0	&	<0.7	  		&	30.9	&		&		&		&	30.5	\\
TIC~58606910	&	0.068	&	4071	$\pm$	122	&	1.93	$\pm$	0.29	&	111	$^{+	240}	_{-	78}$	&	K0 (18)	&	5.9	&	<0.7	  		&	4.7	&	<0.6	  		&	7.5	&	1.2	$\pm$	0.3	&	30.8	&		&		&		&		\\
TIC~61259674	&	0.160	&	10593	$\pm$	73	&	3.21	$\pm$	0.48	&		  	 	  	 	&	B9IVn (13)	&	8.1	&	1.8	$\pm$	0.3	&	12.1	&	1.6	$\pm$	0.2	&	5.0	&	<0.6	  		&	31.0	&		&		&		&	30.8	\\
TIC~68416596	&		&		  		&		  		&		  	 	  	 	&	M0IVe (25)	&	16.7	&	3.4	$\pm$	0.2	&	5.4	&	<0.8	  		&	8.6	&	2.3	$\pm$	0.4	&		&		&		&		&		\\
TIC~86339506	&	0.426	&	4519	$\pm$	80	&	3.29	$\pm$	0.49	&	96	$^{+	219}	_{-	67}$	&	G5 (9)	&	19.7	&	4.6	$\pm$	0.2	&	8.1	&	1.4	$\pm$	0.3	&	20.4	&	9.0	$\pm$	0.2	&	31.4	&	31.6	&		&		&		\\
TIC~114085232	&	0.039	&	4672	$\pm$	150	&	0.34	$\pm$	0.05	&	95	$^{+	211}	_{-	66}$	&	~	&	3.4	&	<0.8	  		&	6.6	&	1.1	$\pm$	0.3	&	3.8	&	<0.6	  		&	30.1	&		&		&		&		\\
TIC~120837437	&	0.206	&	5372	$\pm$	132	&	1.09	$\pm$	0.16	&	81	$^{+	157}	_{-	56}$	&	~	&	7.1	&	0.9	$\pm$	0.2	&	6.8	&	1.0	$\pm$	0.3	&	6.2	&	0.8	$\pm$	0.2	&	31.1	&	31.1	&	30.7	&	30.3	&		\\
TIC~127572976	&	0.084	&	5343	$\pm$	123	&		  		&		  	 	  	 	&	G6III (11)	&	6.1	&	0.8	$\pm$	0.2	&	11.9	&	2.2	$\pm$	0.2	&	4.5	&	<0.7	  		&	31.8	&	32.4	&	31.8	&	31.5	&		\\
TIC~132764448	&	0.058	&	4612	$\pm$	123	&	0.42	$\pm$	0.06	&	95	$^{+	214}	_{-	66}$	&	~	&	3.2	&	<0.8	  		&	3.1	&	<0.9	  		&	7.1	&	1.9	$\pm$	0.4	&		&		&	29.8	&	29.2	&	29.3	\\
TIC~135443051	&	0.045	&	5714	$\pm$	80	&		  		&		  	 	  	 	&	G0V+G5V (7)	&	11.0	&	2.1	$\pm$	0.2	&	6.2	&	0.7	$\pm$	0.1	&	11.2	&	1.6	$\pm$	0.2	&	30.5	&	30.7	&		&		&		\\
TIC~141146667	&		&	3051	$\pm$	157	&	0.16	$\pm$	0.02	&		  	 	  	 	&	~	&	8.4	&	1.8	$\pm$	0.3	&	6.1	&	1.1	$\pm$	0.2	&	6.9	&	1.0	$\pm$	0.2	&		&		&		&		&		\\
TIC~148841573	&	0.288	&	4495	$\pm$	146	&	6.33	$\pm$	0.95	&	117	$^{+	204}	_{-	78}$	&	K7e (28)	&	12.1	&	2.3	$\pm$	0.3	&	4.2	&	<0.8	  		&	15.7	&	2.5	$\pm$	0.2	&	30.8	&	31.0	&		&		&		\\
TIC~150002985	&	0.046	&	3066	$\pm$	122	&		  		&		  	 	  	 	&	M1Ve (2)	&	13.3	&	2.9	$\pm$	0.3	&	3.2	&	<0.5	  		&	10.0	&	1.7	$\pm$	0.3	&		&		&		&		&	29.5	\\
TIC~150173452	&	0.069	&		  		&		  		&		  	 	  	 	&	M0.6 (20)	&	3.3	&	<0.7	  		&	3.5	&	<0.6	  		&	18.3	&	4.3	$\pm$	0.2	&	30.3	&	30.6	&		&		&	29.7	\\
TIC~152346470	&	0.540	&	5844	$\pm$	134	&	1.96	$\pm$	0.29	&	68	$^{+	70}	_{-	47}$	&	G5V (11)	&	3.8	&	<0.6	  		&	8.8	&	1.8	$\pm$	0.3	&	4.0	&	<0.7	  		&	30.8	&	30.7	&	30.8	&	30.4	&		\\
TIC~186777643	&	0.063	&	5796	$\pm$	132	&	3.25	$\pm$	0.49	&	106	$^{+	114}	_{-	72}$	&	G1V (11)	&	15.6	&	2.9	$\pm$	0.3	&	7.6	&	1.4	$\pm$	0.3	&	4.7	&	<0.7	  		&	30.9	&	30.9	&	31.0	&	30.6	&		\\
TIC~195165333$^*$	&	0.355	&	5230	$\pm$	130	&	7.29	$\pm$	1.09	&	371	$^{+	194}	_{-	187}$	&	G9III (11)	&	2.8	&	<0.6	  		&	4.0	&	<0.9	  		&	14.8	&	2.8	$\pm$	0.3	&	31.3	&		&	31.0	&	30.5	&	30.4	\\
TIC~238778709	&	0.148	&	5391	$\pm$	124	&	11.82	$\pm$	1.77	&	1120	$^{+	296}	_{-	285}$	&	F8 (9)	&	6.9	&	2.0	$\pm$	0.5	&	4.1	&	<0.6	  		&	3.1	&	<0.6	  		&	31.1	&	31.2	&		&		&		\\
TIC~241351537	&	0.297	&	4955	$\pm$	145	&		  		&		  	 	  	 	&	G8III+sdO (3)	&	14.9	&	3.6	$\pm$	0.3	&	13.6	&	2.7	$\pm$	0.3	&	5.9	&	<0.8	  		&	32.5	&	32.8	&		&		&		\\
TIC~248024514	&	0.052	&	4970	$\pm$	123	&	11.32	$\pm$	1.7	&	975	$^{+	328}	_{-	364}$	&	G5 (21)	&	9.0	&	1.2	$\pm$	0.2	&	3.4	&	<0.6	  		&	3.8	&	<0.7	  		&		&		&		&		&		\\
TIC~252873509	&	0.122	&	6539	$\pm$	128	&	7.31	$\pm$	1.1	&		  	 	  	 	&	F5 (9)	&	23.0	&	8.3	$\pm$	0.2	&	24.6	&	10.1	$\pm$	0.2	&	22.4	&	5.4	$\pm$	0.2	&	31.7	&	32.0	&		&		&		\\
TIC~268397995	&	0.128	&	3923	$\pm$	63	&	2.72	$\pm$	0.41	&	117	$^{+	248}	_{-	82}$	&	M0-1 (15)	&	3.5	&	<0.8	  		&	3.9	&	<0.6	  		&	10.6	&	1.8	$\pm$	0.2	&		&		&		&		&	30.1	\\
TIC~282780307	&	0.165	&	4693	$\pm$	127	&	6.33	$\pm$	0.95	&	110	$^{+	168}	_{-	74}$	&	~	&	3.3	&	<0.7	  		&	9.2	&	1.2	$\pm$	0.2	&	3.5	&	<0.6	  		&		&	31.4	&		&		&		\\
TIC~285643624	&	0.355	&	4629	$\pm$	127	&	11.06	$\pm$	1.66	&	828	$^{+	403}	_{-	387}$	&	K5 (9)	&	18.1	&	3.7	$\pm$	0.2	&	3.3	&	<0.6	  		&	5.0	&	<0.7	  		&	31.3	&	31.3	&		&		&		\\
TIC~306301020	&	0.245	&	5494	$\pm$	117	&	1.78	$\pm$	0.27	&	78	$^{+	146}	_{-	54}$	&	G2V (27)	&	3.2	&	<0.7	  		&	3.4	&	<0.6	  		&	8.6	&	1.6	$\pm$	0.3	&	30.4	&	30.7	&		&		&		\\
TIC~309965134	&	0.177	&	5107	$\pm$	123	&		  		&		  	 	  	 	&	G5 (21)	&	2.9	&	<0.6	  		&	8.3	&	1.3	$\pm$	0.3	&	5.2	&	<0.6	  		&	30.9	&	31.1	&		&		&		\\
TIC~332487879	&	0.134	&	5192	$\pm$	96	&		  		&		  	 	  	 	&	G5IV (4)	&	3.1	&	<0.7	  		&	16.9	&	3.9	$\pm$	0.3	&	17.9	&	4.4	$\pm$	0.3	&	31.5	&	31.6	&		&		&		\\
TIC~353053003	&		&		  		&	0.37	$\pm$	0.06	&		  	 	  	 	&	M4.5V (23)	&	8.5	&	1.2	$\pm$	0.2	&	9.8	&	2.0	$\pm$	0.3	&	8.4	&	1.6	$\pm$	0.3	&	29.3	&	29.4	&		&		&		\\
TIC~357345626	&	0.215	&	4960	$\pm$	124	&		  		&		  	 	  	 	&	G8III+B9.5e (22)	&	3.9	&	<0.6	  		&	3.8	&	<0.6	  		&	7.7	&	1.6	$\pm$	0.3	&	30.9	&		&		&		&		\\
TIC~364130065	&	0.335	&	5344	$\pm$	131	&		  		&		  	 	  	 	&	K (21)	&	3.0	&	<0.6	  		&	17.3	&	3.6	$\pm$	0.3	&	5.6	&	<0.7	  		&	30.9	&	31.3	&		&		&		\\
TIC~366680372	&	0.172	&	4192	$\pm$	124	&	2.18	$\pm$	0.33	&	106	$^{+	236}	_{-	74}$	&	~	&	10.2	&	1.3	$\pm$	0.2	&	3.4	&	<0.7	  		&	12.8	&	2.5	$\pm$	0.3	&	30.8	&		&		&		&		\\
TIC~389188453	&	0.140	&	5409	$\pm$	122	&		  		&		  	 	  	 	&	G0V (21)	&	3.3	&	<0.7	  		&	3.7	&	<0.6	  		&	8.5	&	1.2	$\pm$	0.3	&	30.7	&	30.9	&		&		&		\\
TIC~399151944	&	0.351	&	4141	$\pm$	125	&	2.07	$\pm$	0.31	&	108	$^{+	238}	_{-	75}$	&	K4IVe (25)	&	19.3	&	4.0	$\pm$	0.3	&	3.4	&	<0.8	  		&	3.3	&	<0.8	  		&	30.6	&	31.0	&	30.4	&	29.9	&		\\
TIC~399181042	&	0.150	&	5618	$\pm$	124	&		  		&		  	 	  	 	&	G5 (9)	&	14.0	&	2.7	$\pm$	0.2	&	3.2	&	<0.6	  		&	6.4	&	0.8	$\pm$	0.2	&	31.3	&	31.4	&	31.2	&	30.7	&		\\
TIC~415030187	&	0.017	&	4583	$\pm$	124	&		  		&		  	 	  	 	&	K0+IIICH-1 (6)	&	3.0	&	<0.7	  		&	4.8	&	<0.7	  		&	19.5	&	5.7	$\pm$	0.3	&	31.6	&	31.7	&		&		&		\\
TIC~432909199$^*$	&	0.252	&	6318	$\pm$	451	&	0.95	$\pm$	0.14	&		  	 	  	 	&	G0 (29)	&	3.6	&	<0.6	  		&	7.2	&	1.1	$\pm$	0.2	&	3.9	&	<0.8	  		&	30.9	&	31.0	&	30.7	&	30.4	&		\\
TIC~436588012	&	0.206	&	6061	$\pm$	509	&		  		&		  	 	  	 	&	G5 (18)	&	3.6	&	<0.7	  		&	8.4	&	1.8	$\pm$	0.3	&	3.8	&	<1.7	  		&	30.9	&	31.3	&	31.1	&	30.6	&		\\
TIC~442932272	&		&	2950	$\pm$	52	&	0.4	$\pm$	0.06	&		  	 	  	 	&	M4.5 (12)	&	4.4	&	<0.9	  		&	5.9	&	<1.0	  		&	12.5	&	4.6	$\pm$	0.5	&	29.1	&	29.0	&	28.9	&	28.2	&		\\
TIC~444062991	&	0.243	&	5227	$\pm$	110	&	0.66	$\pm$	0.1	&	84	$^{+	170}	_{-	58}$	&	G0 (5)	&	4.1	&	<0.6	  		&	3.9	&	<1.0	  		&	8.2	&	1.0	$\pm$	0.2	&	31.1	&	31.6	&		&		&		\\
TIC~444656050	&	0.398	&	5570	$\pm$	152	&	0.78	$\pm$	0.12	&	77	$^{+	142}	_{-	53}$	&	G3IV(e) (11)	&	11.9	&	2.3	$\pm$	0.2	&	3.7	&	<1.0	  		&	5.2	&	<0.7	  		&	30.8	&	31.3	&	30.7	&	30.3	&		\\
TIC~456942208	&	0.481	&	5859	$\pm$	104	&	0.73	$\pm$	0.11	&	73	$^{+	121}	_{-	50}$	&	G8 (18)	&	3.4	&	<0.9	  		&	8.0	&	1.2	$\pm$	0.2	&	3.8	&	<0.8	  		&	30.4	&	31.2	&		&		&		\\

 \end{longtable}
 {\textit{Notes:} Column 1 -- \textit{TESS} Input catalogue number of the star, Column 2 -- flare rate (flares per day above $10^{34}$~erg), Column 3 -- rotation period, Column 4 -- gyrochronological age estimate, Column 5 -- effective temperature, Column 6 -- spectral type (1) \citet{1949ApJ...109..231J},(2) \citet{1977ApJ...214..747H},(3) \citet{1982MNRAS.201..901D},(4) \citet{1982MSS...C03....0H},(5) \citet{1988A&AS...74..449R},(6) \citet{1989ApJS...71..245K},(7) \citet{1990A&A...230..389S},(8) \citet{1993AJ....105.1519P},(9) \citet{1993yCat.3135....0C},(10) \citet{1995A&AS..110..367N},(11) \citet{2006A&A...460..695T},(12) \citet{2006AJ....132..866R},(13) \citet{2007A&A...463..671R},(14) \citet{2007AJ....133.2524W},(15) \citet{2009A&ARv..17..251S},(16) \citet{2012A&A...538L...3R},(17) \citet{2012A&A...542A.115B},(18) \citet{2012ApJ...745..119N},(19) \citet{2012ApJS..202....7X},(20) \citet{2014ApJ...786...97H},(21) \citet{2014ASPC..485..223B},(22) \citet{2015A&A...573A.107H},(23) \citet{2015A&A...577A.128A},(24) \citet{2015yCat.5145....0M},(25) \citet{2016MNRAS.461..794P},(26) \citet{2017ApJ...838..150K},(27) \citet{2018A&A...612A..96F},(28) \citet{2019ApJ...877...60B},(29) \citet{2021ApJS..254...20L}, Columns 7,9,11 -- signal-to-noise ratio of the epoch 1--3 VLASS maps, Columns 8,10,12 -- flux density at 3~GHz in the VLASS 1--3 epoch observations. Upper limits for the non-detections are given at  $5\sigma$ rms noise levels, Columns 13--17 -- $\log$ X-ray luminosities from the first \citep{1999A&A...349..389V,2000IAUC.7432....3V} and second \citep{2016A&A...588A.103B} \textit{ROSAT}, and the \textit{SRG/eROSITA} \citep{2024A&A...682A..34M} all-sky surveys, and the XMM-Newton serendipitous survey \citep{2016A&A...590A...1R}.}
\end{landscape}
\FloatBarrier
\clearpage

\end{appendix}
\end{document}